%% file: SSIP.tex
\documentclass[a4paper, onecolumn, 11pt]{quantumarticle}
\pdfoutput=1
\usepackage{breakurl}
\usepackage{stmaryrd}
\usepackage{amsthm,amstext,amssymb}
\usepackage{epsfig}
\usepackage{slashed}
\usepackage{fancyhdr}
\usepackage{graphicx}
\usepackage{pdfpages}
\usepackage{float}
\usepackage{subfig}
\usepackage{simplewick}
\usepackage{graphics,psfrag}
\usepackage{pstricks}
\usepackage{latexsym}
\usepackage{changepage}
\usepackage{caption}
\usepackage{paralist}
\usepackage{tikzit}
\input{basic.tikzstyles}
\input{quantum.tikzstyles}
\usepackage{graphicx}
\usepackage{array}
\usepackage{hyperref}
\usepackage{tikz-cd}
\usepackage{mathrsfs}
\usepackage{braket}
\usepackage{soul}

\usepackage{hhline}
\usepackage{algorithm}
\usepackage{algpseudocode}

\usepackage{scalefnt}
\usepackage[all]{xypic}
\usepackage{slashed}
\usepackage{extarrows}
\usepackage{lettrine}
\usepackage{booktabs}
\usepackage{paralist}
\usepackage{subfig}
\usepackage{wrapfig}
\usepackage{xcolor}

\DeclareMathAlphabet{\mathcal}{OMS}{cmsy}{m}{n}
\newtheorem{lemma}{Lemma}[section] 

\newtheorem{definition}[lemma]{Definition}
\newtheorem{remark}[lemma]{Remark}

\renewcommand{\imath}{\mathrm{i}}

\newcommand{\CG}{\hbox{{$\mathcal G$}}}

\newcommand{\CL}{\hbox{{$\mathcal L$}}}

\newcommand{\CP}{\hbox{{$\mathcal P$}}}

\newcommand{\CO}{\hbox{{$\mathcal O$}}}

\newcommand{\R}{\mathbb{R}}
\newcommand{\F}{\mathbb{F}}

\newcommand{\del}{\partial}

\newcommand{\tens}{\mathop{{\otimes}}}

\newcommand{\id}{\mathrm{id}}
\newcommand{\im}{\mathrm{im}}

\newcommand{\MatF}{\mathtt{Mat}_{\F_2}}
\newcommand{\Chains}{\mathtt{Ch}(\MatF)}

\newsavebox{\pullback}
\sbox\pullback{%
\begin{tikzpicture}%
\draw (0,0) -- (2ex,0ex);%
\draw (2ex,0ex) -- (2ex,2ex);%
\end{tikzpicture}}

\newsavebox{\pushout}
\sbox\pushout{%
\begin{tikzpicture}%
\draw (0,0) -- (0,2ex);%
\draw (0,2ex) -- (2ex,2ex);%
\end{tikzpicture}}

\setstcolor{red}

\begin{document}
\title{SSIP: automated surgery with quantum LDPC codes}

\author{Alexander Cowtan}
\affiliation{Dept. of Computer Science,  University of Oxford,
Wolfson Building, Parks Road, Oxford OX1 3QD, UK}
\affiliation{Quantinuum,
Terrington House, 13-15 Hills Road, Cambridge CB2 1NL, UK}
\email{akcowtan@gmail.com}

\maketitle

\begin{abstract}
We present Safe Surgery by Identifying Pushouts (\verb|SSIP|), an open-source lightweight Python package for automating surgery between qubit CSS codes. \verb|SSIP| is flexible: it is capable of performing both external surgery, that is surgery between two codeblocks, and internal surgery, that is surgery within the same codeblock. Under the hood, it performs linear algebra over $\F_2$ governed by universal constructions in the category of chain complexes. We demonstrate on quantum Low-Density Parity Check (qLDPC) codes, which are not topological codes in general, and are of interest for near-term fault-tolerant quantum computing. Such qLDPC codes include lift-connected surface codes, generalised bicycle codes and bivariate bicycle codes. We show that various logical measurements can be performed cheaply by surgery without sacrificing the high code distance. For example, half of the single-qubit logical measurements in the $Z$ or $X$ basis on the $\llbracket 144 ,12, 12 \rrbracket$ gross code require only 30 total additional qubits each, assuming the upper bound on distance given by \verb|QDistRnd| is tight. This is two orders of magnitude lower than the additional qubit count of 1380 initially predicted by Bravyi et al. \cite{BCGMRY}.
\end{abstract}

\section{Introduction}
Substantial progress has recently been made in designing quantum error-correcting codes with favourable properties, both in the asymptotic regime \cite{PK1, BE1, PK2, LZ, DHLV} and at finite blocklength \cite{PK2, PK3, BCGMRY, KoPr, LP, SHR}. However, it is still largely unknown how best to perform universal computation while maintaining fault-tolerance. There are some proposals for protocols for logical operations \cite{Kris,Coh,Burt1,QWV,HJY,BDB,ES,ZSPCB}, which typically depend on the precise nature of the code used as a quantum memory.

There are several desiderata for logical operations on codes. They should:
\begin{itemize}
\item	Yield universality – commonly in conjunction with state injection.
\item	Be individually addressable on logical qubits.
\item	Be parallelisable.
\item	Not add significant overhead to the quantum memory – in terms of qubit count, stabiliser weight, reduction in threshold etc.
\end{itemize}

Logical operations can be performed with surface codes, and other topological codes, by lattice surgery \cite{HFDM, FMMC}, a procedure which satisfies all of the above desiderata, including universality when accompanied by magic state distillation \cite{BK}. Unfortunately, surface codes are restricted in their encoding rate and distance, and we would prefer a quantum memory with better parameters. We would like to eschew topological codes, while keeping the ability to do surgery. Lattice surgery can be generalised to Calderbank-Shor-Steane (CSS) codes which are not topological \cite{Coh, CowBu}. This is somewhat subtle, and is harder to reason about than the topological case.

Here we present \verb|SSIP|, software which automates the procedure of identifying and performing CSS code surgery. While we focus on the homological formalism in \cite{CowBu}, the software is also capable of performing some of the surgeries in \cite{Coh} by converting the protocols defined using Tanner graphs into chain complexes. \verb|SSIP| has been extensively tested and benchmarked, and we find that it is fast (and correct) on small-to-medium sized codes, while using lower resource requirements than previously estimated \cite{BCGMRY}.

The layout of this paper is as follows. We start by giving a brief recap of the well-known CSS code-homology correspondence \cite{BM, BE2}, then explain how \verb|SSIP| determines if two logical operators from different codeblocks can be merged together, yielding a logical parity measurement. This is an external code merge. \verb|SSIP| does not explicitly handle code splits, the adjoint operation to merges, because once a merge has been found we have all the data required for its corresponding split. Upon performing a merge, \verb|SSIP| can optionally compute substantial additional data, such as which new ancillae data qubits, stabilisers and logical qubits are introduced. We illustrate first with some very small codes, including mildly interesting cases where we merge a triorthogonal code into other quantum memories, allowing for magic state injection without distillation. All examples can be found in the Github repository \verb|https://github.com/CQCL/SSIP|. We then give results for a variety of external merges with lift-connected surface codes \cite{ORM}, generalised bicycle codes \cite{KoPr}, and bivariate bicycle codes \cite{BCGMRY}.

We perform logical single-qubit and parity measurements in the $X$ and $Z$ bases. After a merge is performed, we must ensure that the code distance is preserved; for small codes this is straightforward to calculate using naive methods, such as enumerating over all logical operators, but for codes with blocklengths in the hundreds of qubits such methods would take too long. We use \verb|QDistRnd| \cite{QDR} to upper bound the code distances. Where possible we use the Satisfiability Modulo Theories (SMT) solver Z3 \cite{MB} to give explicit distances.

We demonstrate the developed techniques on the $\llbracket 144, 12, 12\rrbracket$ gross code from \cite{BCGMRY}, which belongs to a family of bivariate bicycle codes. We find that we can measure any logical qubit simultaneously in either the $X$ or $Z$ basis while maintaining $d = 12$, when viewed as a subsystem code, using at most an additional 78 data qubits, and 72 syndrome qubits, so 150 total ancillae. This is substantially lower than the $1380$ additional qubits described in \cite[Sec.~9.4]{BCGMRY}. These results on the gross code are reliant on the upper bound from \verb|QDistRnd| being tight.

We then describe how one can perform internal merges within a CSS codeblock. Happily, the procedure is very similar to external merges, which we previously described in \cite{CowBu}, with almost the same prerequisites.
We find that we can perform pairwise parity measurements between many (but not all) of the logical qubits in the gross code in either basis using a total of at most 150 extra qubits, maintaining code distance.

Importantly, there are several things which \verb|SSIP| does \textit{not} do. For fault-tolerance, we must give circuits for syndrome measurements and other operations on the codes, and then use an accurate error model to establish a (pseudo-)threshold \cite{KLZ}. Without circuits, one can attempt to approximate the error tolerance of the code using phenomenological noise. \verb|SSIP| does neither of these things; it does not include methods for constructing quantum circuits or modelling noise in any way. Additionally, codes should come with good decoders, allowing us to extract a likely error from the outcomes of syndrome measurements \cite{PK3, WB, RWBC}. \verb|SSIP| does not perform any decoding. Lastly, quantum architectures commonly have geometric constraints, which put conditions on the Tanner graphs of any implemented CSS codes. \verb|SSIP| has no notion of geometric constraints or architectures. For simplicity, it is solely concerned with code parameters and figures of merit, such as code distances and stabiliser weights.

\subsection{Contributions}
In this paper we add some new algebraic theory to the work in \cite{CowBu}, concerning the `depth' of merges and formulating internal merges of codes. We also translate some of the Tanner graph formalism from \cite{Coh} into chain complexes, which may be of independent interest. Algebraic theory is not the primary aim of our work, however, and the bulk of our contributions are in the software and the results thereby obtained.

\subsection{Related work}\label{ref:related}
There is a sizeable body of work on performing transversal gates with qLDPC codes \cite{Burt1, ES, QWV, ZSPCB}, but due to the no-go theorem of Eastin-Knill \cite{EK} these cannot be sufficient for universality and must be augmented with logical operations which are not transversal.

Cohen et al \cite{Coh} first published generalisations of lattice surgery to arbitrary CSS codes. Our work is directly inspired by theirs, although our approach is homological rather than using Tanner graphs. Our surgeries, when performing logical parity measurements and their adjoint splits, are also different. This makes performing an apples-to-apples comparison between the two difficult, but we do present some comparisons to their approach.

There are several different open-source repositories for reasoning about quantum CSS and LDPC codes \cite{Sabo,Perl,Roffe}. These have different foci, and to the best of our knowledge none are designed to reason about surgery.

Separately, LaSsynth has recently been developed for synthesising lattice surgeries \cite{TNG}. This has a different scope to the present work. LaSsynth takes a desired quantum routine and synthesises it into a sequence of lattice surgery operations, encoding the synthesis problem as a SAT instance to exhaustively optimise the resources. It is designed exclusively for surface codes.  In a similar vein see \cite{Wat}. \verb|SSIP| cannot compile quantum routines, beyond a specified merge/split or sequence of merges/splits. It could be fruitful to attempt to optimise resources when performing surgery with more elaborate codes than surface codes in a similar manner.

\subsection{General software description}

The software is called Safe Surgery by Identifying Pushouts (\verb|SSIP|) because its core function is to find pushouts, and other colimits, between codes in order to perform surgery. This surgery is `safe' in the sense that it is guaranteed to perform logical measurements on the logical qubits involved in the merge, without affecting logical data elsewhere in the code. It is also safe in the sense that the distance can be checked afterwards, although this becomes challenging for codes at high blocklengths and distances. As a consequence, \verb|SSIP| is best suited to codes with blocklengths in the low hundreds, i.e. for near-term fault-tolerant computing.

\verb|SSIP| is written in Python for ease of use, with occasional function calls to a library in GAP \cite{QDR} for code distance estimates. \verb|SSIP| is available from its Github repository \verb|https://github.com/CQCL/SSIP| or alternatively by calling \verb|pip install ssip|, and is fully open-source, released with a permissive MIT license. Documentation can be found at \verb|https://cqcl.github.io/SSIP/api-docs/|.

For simplicity, \verb|SSIP| is entirely procedural. The only new classes defined in \verb|SSIP| are structs\footnote{Python does not have structs, but as of Python 3.7 it has dataclasses, which are close to structs.}, such as the \verb|CSScode|, which merely contains the two parity-check matrices of a CSS code, stored as \verb|numpy| arrays. The codebase then operates by performing numerics in functions, passing around the \verb|CSScode| and other elementary data structures.

There are four main purposes of \verb|SSIP|:
\begin{itemize}
\item Determine whether, given suitable data, a code merge is possible (and hence its adjoint split).
\item Perform code merges and return merged codes.
\item Calculate additional data about the merge, such as the new stabilisers, data qubits and logical qubits introduced.
\item Calculate code parameters, either as CSS codes or as subsystem codes, once merges have been performed.
\end{itemize}

All of these are extremely tedious to compute by hand, and so software is required for practically relevant codes.

We will describe how \verb|SSIP| performs all of these steps, but in order to explain our algorithms and results we must give some algebraic background on CSS codes and surgery.

\section{The CSS code-homology correspondence}

We give a summary of the aspects of CSS codes and homology required to understand this paper. For a more in-depth description of surgery via homology see \cite{CowBu}. For more on the CSS code-homology correspondence see \cite{BE2} and the references therein. For the basics of homology, which are sufficient to understand our work, see \cite[Ch. 1-2]{Weib}.

The standard definition of a qubit CSS code is a pair of orthogonal binary linear codes of the same length $\mathcal{C}_Z, \mathcal{C}_X \subset \F_2^n$ such that $\mathcal{C}_Z \subset\mathcal{C}_X^\perp$ \cite{CS}; this also implies $\mathcal{C}_X \subset \mathcal{C}_Z^\perp$. At this level of abstraction, $\mathcal{C}_Z$ and $\mathcal{C}_X$ are merely subspaces. We can always equip $\mathcal{C}_Z$ with a parity-check matrix $P_Z: \F_2^n \rightarrow \F_2^{m_Z}$ such that $\mathcal{C}_Z = \ker(P_Z)$, and the same for $\mathcal{C}_X$. Then, the condition that $\mathcal{C}_Z \subset \mathcal{C}_X^\perp$ is equivalent to $\im (P_Z^\intercal) \subset \ker (P_X)$, and vice versa.

Thus we have a chain complex 
\[C_\bullet = \begin{tikzcd}\F_2^{m_Z} \arrow[r, "\del_2"] & \F_2^n \arrow[r, "\del_1"] & \F_2^{m_X}\end{tikzcd},\] 
where by convention $\del_2 = P_Z^\intercal, \del_1 = P_X$. We set $C_2 = \F_2^{m_Z}$, $C_1 = \F_2^n$, $C_0 = \F_2^{m_X}$. By duality we also have a cochain complex 
\[C^\bullet = \begin{tikzcd}\F_2^{m_X} \arrow[r, "\delta^0"] & \F_2^n \arrow[r, "\delta^1"] & \F_2^{m_Z}\end{tikzcd},\] 
with $\delta^0 = P_X^\intercal$, $\delta^1 = P_Z$, setting $C^0 = C_0$ etc.

There is more data contained in the (co)chain complexes than merely the CSS code as a stabiliser group and its stabilised subspace. In choosing the matrices $P_Z, P_X$ we have also assigned generators to the stabiliser group, which will be syndrome measurements on the code. Rows of $P_Z(P_X)$ are generators as tensor products of Paulis, with $0$ representing $I$ and $1$ representing $Z(X)$.

It is straightforward to check that, given $u\in \ker(P_X)$, elements of the equivalence class $u + \im(P_Z^\intercal)$ are equivalent $\overline{Z}$ logical operators on the codespace, and elements in $[0] = \im(P_Z^\intercal)$ act trivially on the codespace.

Let $H_1(C_\bullet)$ be the first homology space of $C_\bullet$, i.e. $H_1(C_\bullet) = \ker(P_X)/\im(P_Z^\intercal)$. Then $\dim H_1(C_\bullet) = k$, where $k$ is the number of logical qubits, and a choice of basis for $\dim H_1(C_\bullet)$ is a choice of tensor decomposition of the logical space into logical qubits. The same applies to the cohomology space $H^1(C^\bullet)$. This tensor decomposition is consistent when the symmetric bilinear nondegenerate form, i.e. the dot product
\[\cdot : \F_2^n \times \F_2^n \rightarrow \F_2;\quad u\cdot v = u^\intercal v\]
yields $[u]_i \cdot [v]_j = \delta_{i,j}$ for equivalence classes $\{[u]_i\}_{i\in I}$ in the basis set $I$ of $H_1(C_\bullet)$ and $\{[v]_j\}_{j\in J}$ in the basis set $J$ of $H^1(C^\bullet)$. That is, $\overline{Z}_i$ anticommutes with $\overline{X}_i$ and commutes with $\overline{X}_j$ on any other logical qubit.

The lowest weight nontrivial vector $u \in \ker(P_X)\backslash \im(P_Z^\intercal)$ is the nontrivial $\overline{Z}$ operator on the stabiliser codespace with the lowest number of data qubits in its support; similar for $v \in \ker(P_Z)\backslash \im(P_X^\intercal)$. Let $d_Z = |u|$, $d_X = |v|$, where $|\cdot |$ is the Hamming weight, and then the code distance $d = \min(d_Z, d_X)$. Overall a CSS code is said to have parameters $\llbracket n, k, d \rrbracket$ where $n$ is the number of data qubits, $k$ the number of logical qubits and $d$ the code distance.

A CSS code is called \textit{$\omega$-limited} when the weights of rows and columns in both parity-check matrices are bounded above by $\omega$. It is common to consider infinite families of codes of increasing size. If every member of the family is $\omega$-limited for some finite $\omega$ then the family is called quantum Low-Density Parity Check (qLDPC). A similar definition applies to classical LDPC codes.

Throughout, we will refer to the $\omega$ of a code as being the maximum column or row weight of its parity-check matrices.

\begin{remark}
In order for the definitions to do with weights to be sensible we require the chain complexes to be precisely of the form $\F_2^{m_Z} \rightarrow \F_2^n \rightarrow \F_2^{m_X}$ on the nose, not just up to isomorphism on the components; they must also have basis elements being entries in the vector spaces.
\end{remark}

A subsystem CSS code is a CSS code where some of the logical qubits are not used for storing logical data. These qubits are instead called gauge qubits \cite{KLP}. In this case the first homology space divides into $H_1(C_\bullet) = \CL \oplus \CG$, with $\CL$ being the space used for storing data and $\CG$ the space of gauge qubits.

Importantly, we can relegate logical qubits to be gauge qubits instead in order to increase the distance of the code. The nontrivial $\overline{Z}$ logicals which act on the logical data must have some support in $\CL$, not just $\CG$. These belong to the set $H_1(C_\bullet)\backslash \CG$ and are called dressed logical operators. Dressed logicals can still have support in $\CG$. The same applies to $H^1(C^\bullet)$.

The distance $d$ of a subsystem CSS code is therefore the smallest weight of the dressed logical operators.

\subsection{Code distance}
We take a brief aside to discuss the calculation of minimum distance for CSS codes. There are at least 5 ways to perform this calculation for CSS codes, without relying on the codes being 2D surface codes using e.g. \cite[App.~B]{Br}.

\begin{enumerate}
\item Enumerate over all logicals in the code and find the one(s) with the lowest weight. The compute time of this will generally scale exponentially in $n$, and so it can only reasonably be used for small codes. Evidently the compute time is insensitive to $d$, as every operator is checked regardless.

\item Start by searching for any weight 1 logicals and then increment the weight until a logical is found. This will also generally scale poorly but will perform better for codes with low $d$, even if $n$ is high.

\item Use \verb|QDistRnd| to give an upper bound on the code distance \cite{QDR}. To calculate $d_Z$ \verb|QDistRnd| constructs a generator matrix whose rows are a basis of $\ker(P_X)$, then randomly permutes columns, performs Gaussian elimination, and un-permutes the columns, leaving a random set of rows in $\ker(P_X)$. Rows not in $\im(P_Z^\intercal)$ are then considered for their lowest weight. The permutation is applied many times, improving the upper bound on $d_Z$. The same can then be done for $d_X$.

\item Interpret the minimum distance problem as a binary programming problem. This is a somewhat less common problem than the mixed integer programming problem, so in certain cases one can convert the former into the latter to make use of mixed integer programming solvers \cite[Sec.~C 1.]{LAR}.

\item Perform distance verification with ensembles of codes with related properties \cite{DKP}.
\end{enumerate}

\verb|SSIP| makes use of the first three methods. In our results, for small codes or codes for which we already know the distance is modest, we use (1.) and (2.). For some codes we upper bound the distance using (3.) first and then verify that the bound is tight using (2.). In \verb|SSIP| we offload the computation of finding logicals in (2.) to Z3 \cite{MB}, which is written in C++ and so significantly faster than it would be to find logicals in Python. For the largest codes, in the hundreds of qubits, we merely estimate the distance using (3.). While there are no guarantees, we find that empirically \verb|QDistRnd| is accurate compared to exact results given a large enough number of information sets, say $10^4$ for codes in the low hundreds of qubits.

We will also calculate the distance of subsystem CSS codes. For computations with methods (1.) and (2.) nothing much changes, we just add conditions to the logicals to consider. For method (3.), we check in Appendix~\ref{app:sub_distance} that any black box method for calculating the code distance of a CSS code can be adapted to subsystem CSS codes, and so by changing the input given to \verb|QDistRnd| we can also use (3.) to upper bound subsystem CSS code distances.

We do not use methods (4.) and (5.) in our results. To the best of our knowledge, the conversion in (4.) to mixed integer programming requires the codes to have regular stabiliser weights, which merged codes will generally not have. The methods in (5.) also require the codes to have a certain structure.

\subsection{Products}
\subsubsection{Tensor products}\label{sec:tensor_recap}
Here we recap tensor products \cite{AC} of classical codes from the perspective of homological algebra. Let $C_\bullet$ and $D_\bullet$ be two linear binary codes 
\[\begin{tikzcd}C_\bullet = C_1 \arrow[r, "A"] & C_0\end{tikzcd};\quad \begin{tikzcd} D_\bullet = D_1 \arrow[r, "B"] & D_0\end{tikzcd}\]
where $A$ and $B$ are parity matrices. The dual codes $C^\bullet$, $D^\bullet$ are the codes obtained by transposing the parity-check matrices. The codes have parameters $[n_A, k_A, d_A]$ and $[n_B, k_B, d_B]$, and their dual codes have parameters $[n_A^\intercal, k_A^\intercal, d_A^\intercal]$ and $[n_B^\intercal, k_B^\intercal, d_B^\intercal]$. Explicitly, 
\[n_A = \dim C_1 ;\quad k_A = \dim \ker(A) ;\quad n_A^\intercal = \dim C_0 ;\quad k_A^\intercal = \dim \ker(A^\intercal)  = \dim C_0/\im(A),\] so $n_A, k_A$ are the length and dimension of the code $C_\bullet$, and $n_A^\intercal, k_A^\intercal$ are the length and dimension of the dual code $C^\bullet$. We will also use the respective distances $d_A, d_A^\intercal$. Similar definitions apply to $B$.

The tensor product quantum code is given by the chain complex:

\[\begin{tikzcd}(C\tens D)_\bullet = C_1\otimes D_1 \arrow[r, "\del_2"] & C_0 \otimes D_1 \oplus C_1 \otimes D_0 \arrow[r, "\del_1"] & C_0\otimes D_0 \end{tikzcd} \]
where by convention we say that $\del_2 = P_Z^\intercal$ and $\del_1 = P_X$, and
\[P_Z = \begin{pmatrix}A^\intercal \tens \id_{D_1} &\id_{C_1} \tens B^\intercal\end{pmatrix}; \quad P_X= \begin{pmatrix} \id_{C_0} \tens B & A \tens \id_{D_0}\end{pmatrix}\]

We will use some straightforward facts about this code. It has parameters 
\[\llbracket n_A^\intercal n_B + n_An_B^\intercal, k_A^\intercal k_B+ k_Ak_B^\intercal, \min(d_A,d_B,d_A^\intercal,d_B^\intercal)\rrbracket.\]
That $n_{(C\tens D)} = n_A^\intercal n_B + n_An_B^\intercal$ is obvious from $\dim (C\tens D)_1$. $k_{(C\tens D)} = \dim H_1((C\tens D)_\bullet) = \dim \ker(P_X)/\im(P_Z^\intercal) $, which can be found using the K{\"u}nneth formula \cite{Weib}:
\[H_1((C\tens D)_\bullet) \cong H_0(C_\bullet) \tens H_1(D_\bullet) \oplus H_1(C_\bullet) \tens H_0(D_\bullet) \]
In particular we have the decomposition
\begin{equation}\ker(P_X) = \ker(P_X)/\im(P_Z^\intercal) \oplus \im(P_Z^\intercal) = C_0/\im(A) \otimes \ker(B) \oplus \ker(A)\otimes D_0/\im(B) \oplus \im(P_Z^\intercal).\end{equation}\label{eq:ker_decomp}
The distance can be obtained easily using this expression for $\ker(P_X)$ and a similar one for $\ker(P_Z)$:
\begin{equation}\ker(P_Z) = \ker(A^\intercal) \otimes D_1/\im(B^\intercal) \oplus C_1/\im(A^\intercal) \otimes \ker(B^\intercal) \oplus \im(P_X^\intercal)\end{equation}
These equations are only required by the K{\"u}nneth formula to hold up to isomorphism, but one can check using a simple counting argument that they hold on the nose.

Considering families of tensor product codes built from families of classical codes, the products will be qLDPC iff the classical codes are also LDPC, and the parameters above mean that the scaling have $d \in \CO(\sqrt{n})$ and $k \in \CO(n)$. These bounds are saturated by quantum expander codes \cite{LTZ}, using the fact that hypergraph product codes \cite{TZ} are tensor product codes with one of the classical codes dualised.

Tensor product codes are important to us because we will use them to `glue' other codes together, and for performing single-qubit logical measurements.

\subsubsection{Lifted products}\label{sec:lifted_prods}

Lifted products are a mild generalisation of tensor products, and we will make use of lifted products extensively in our set of examples. Unlike tensor products, they are not a necessary ingredient of our constructions, but many lifted product codes have good parameters -- in both the formal and informal senses \cite{PK1,PK2,PK3,KoPr,ORM,BCGMRY,LP,SHR} -- and so are a useful class of codes on which to demonstrate our methods.

Recall that a chain complex is well-defined over any ring $R$. Differentials are $R$-module homomorphisms. We assume that the components of the chain complexes are all free $R$-modules of finite rank.

Then, fix $R$ to be a commutative subring of $\mathcal{M}_{\ell}(\F_2)$, the ring of $\ell$-by-$\ell$ matrices over $\F_2$, with a specified basis. The tensor product of chain complexes is also well-defined for the ring $R$. Taking two chain complexes over $R$ and making the tensor product $(C \underset{R}{\otimes} D)_\bullet$, this tensor product is also a valid chain complex when replacing each entry of the differentials in $R$ with its corresponding matrix over $\F_2$, and considering the whole chain complex over $\F_2$. This is the lifted product.

Given two initial classical codes $C_\bullet$ and $D_\bullet$, which are length 1 chain complexes over $R$, and so also over $\F_2$, we do not generally know \textit{a priori} what the code parameters $k$ and $d$ of the lifted product code $(C \underset{R}{\otimes} D)_\bullet$ will be when viewed over $\F_2$, a marked difference from the tensor product. The straightforward facts derived from the K{\"u}nneth formula only apply to the complex viewed over $R$, and do not easily translate to $\F_2$.

When $R = \F_2$, the lifted product coincides with the tensor product. Lifted products are special cases of balanced products \cite{BE2} where the actions are free. A common ring to use for generating codes is $\mathscr{C}_\ell$, the ring of $\ell$-by-$\ell$ circulant matrices. This is guaranteed to be commutative, so the tensor product is defined. Helpfully, $\mathscr{C}_{\ell} \cong \F_2^{\langle \ell \rangle}$, where $\F_2^{\langle \ell \rangle} := \F_2[x]/(x^\ell - 1)$ is the ring of polynomials over $\F_2$ modulo $x^\ell -1$, by sending the $m$th shift matrix to $x^m$. This means that any circulant matrix can be denoted concisely by its corresponding polynomial.

\verb|SSIP| can generate a variety of different lifted products, but there are several families of lifted product codes which one would have to generate elsewhere and import. Given that a \verb|CSScode| in \verb|SSIP| is just a pair of \verb|numpy| arrays, this is straightforward.

\subsection{CSS code surgery}

We can consider not only individual (co)chain complexes, but also the category of (co)chain complexes. See \cite{Lein} for a basic primer on category theory. We restrict to chain complexes, with the understanding that everything translates through to cochain complexes by duality.

The category $\Chains$ has as objects chain complexes over $\F_2$. Morphisms are chain maps, matrices between components at the same degree, such that the matrices are coherent in the sense that we have the following commuting squares:
\[\begin{tikzcd}\cdots \arrow[r] & C_{n+1}\arrow[r, "\del^{C_\bullet}_{n+1}"]\arrow[d, "f_{n+1}"] & C_{n}\arrow[r, "\del^{C_\bullet}_{n}"]\arrow[d, "f_{n}"] & C_{n-1}\arrow[r]\arrow[d,"f_{n-1}"] & \cdots\\
\cdots \arrow[r] & D_{n+1}\arrow[r, "\del^{D_\bullet}_{n+1}"] & D_{n}\arrow[r, "\del^{D_\bullet}_{n}"] & D_{n-1}\arrow[r] & \cdots\end{tikzcd}
\]

\begin{definition}(Basis-preserving)
We say that a chain map is basis-preserving when each matrix sends basis elements to basis elements, i.e. they are functions on basis elements.
\end{definition}

We can use universal properties in $\Chains$ to construct new codes from old ones; in particular, we can perform surgery between CSS codes by using pushouts and coequalisers. We give a quick recap of this procedure here. See \cite{CowBu} for a more detailed explanation.

We assume that we are performing $\overline{Z}$-parity measurements, which merge codes in a manner which uses $\Chains$; $\overline{X}$-parity measurements can be inferred by duality, using the category of cochain complexes instead. Similarly, splits of codes can be inferred by reversing the procedure.

\begin{definition}
Let $v \in \ker(P_X)\backslash \im(P_Z^\intercal)$ be such that no other vector in $\ker(P_X)$ is contained in the support of $v$. Then we call $v$ an irreducible logical operator.
\end{definition}

\begin{remark}
We previously called ``irreducible" logical operators ``separated" in \cite{CowBu}. We have adopted the term ``irreducible" for consistency with the literature \cite{DKP, DKP2}.
\end{remark}

\begin{definition}(Logical operator subcomplex)\label{def:log_op_subcomplex}
Given a logical $\overline{Z}$ operator $v \in \ker(P_X)\backslash\im(P_Z^\intercal)$, we can construct a chain complex which represents this operator and its stabilisers, in a suitable sense.

$V_\bullet = V_1 \rightarrow V_0$, where:
\[\tilde{V}_1 = {\rm supp\ } v; \quad \del^{V_\bullet}_{1} = \del^{C_\bullet}_{1}\restriction_{{\rm supp\ } v}; \quad \tilde{V_{0}} = \bigcup_{u \in \im(\del^{V_\bullet}_{1})} {\rm supp\ } u \]
where ${\rm supp\ } v$ is the set of basis vectors in the support of $v$, and $\del_i\restriction_S$ is the restriction of a differential to a subset $S$ of its domain. $V_\bullet$ is called a logical operator subcomplex.
\end{definition}

We have a suitable dualised definition for a logical $\overline{X}$ operator in $\ker(P_Z)\backslash\im(P_X^\intercal)$. We will make repeated use of logical operator subcomplexes throughout. Observe that this subcomplex only has one non-zero differential, i.e. it can be considered a classical code.

\subsubsection{External merges}\label{sec:ext_merges}

Given an irreducible logical operator $v$ in a code we can construct its logical operator subcomplex $V_\bullet$. Then, if we have a monic span:
\[\begin{tikzcd}
V_\bullet \arrow[r, hookrightarrow, "f_\bullet"]\arrow[d, hookrightarrow, "g_\bullet"'] & C_\bullet\\
D_\bullet & &
\end{tikzcd}\]
where both chain maps are basis-preserving, and $V_\bullet$ is a logical operator subcomplex in both codes $C_\bullet$ and $D_\bullet$, then the logical operator is `present' in both codes in a suitable sense, and we can perform an external merge which performs a parity measurement on the two logical qubits.

We do this by first generating a new tensor product code. 

\begin{definition}
Let $P_\bullet = \F_2^r \rightarrow \F_2^{r+1}$ be the classical code with parity-check matrix
\[\del_1^P = \begin{pmatrix}
1 & 0 & 0 & \cdots & 0 \\
1 & 1 & 0 & \cdots & 0 \\
0 & 1 & 1 & \cdots & 0 \\
0 & 0 & 1 & \cdots & 0 \\
\vdots & \vdots & \vdots & \ddots & \vdots \\
0 & 0 & 0 & 1 & 1 \\
0 & 0 & 0 & 0 & 1 \\
\end{pmatrix}\]
i.e. $P_\bullet$ is the incidence matrix of the path graph $\CP_{r+1}$. We call $r$ the depth.
\end{definition}

We can see that $\dim \ker(\del_1^P) = 0$, and $\dim \ker((\del_1^P)^\intercal) = 1$, with the non-zero codeword $1_{r+1}$.

Let $W_\bullet = (P\otimes V)_\bullet$ be the new tensor product code. Explicitly it is the chain complex
\[\F_2^r \otimes V_1 \rightarrow \F_2^{r+1}\otimes V_1 \oplus \F_2^r \otimes V_0 \rightarrow \F_2^{r+1}\otimes V_0\]
with differentials
\[\del^W_2 = \begin{pmatrix}
\id_{V_1} & 0 & 0 & \cdots \\
\id_{V_1} & \id_{V_1} & 0 & \cdots \\
0 & \id_{V_1} & \id_{V_1} & \cdots \\
0 & 0 & \id_{V_1} & \cdots \\
\vdots & \vdots & \vdots & \ddots \\ 
\del_1^V & 0 & 0 & \cdots \\
0 & \del_1^V & 0  & \cdots \\
0 & 0 & \del_1^V  & \cdots  \\
\vdots & \vdots & \vdots & \ddots \\
\end{pmatrix},\quad 
\del^W_1 = \begin{pmatrix}
\del_1^V & 0 & 0 & 0 & \cdots & \id_{V_0} & 0 & 0 & \cdots \\
0 & \del_1^V & 0 & 0 & \cdots & \id_{V_0} & \id_{V_0} & 0 & \cdots \\
0 & 0 & \del_1^V & 0 & \cdots & 0 & \id_{V_0} & \id_{V_0} & \cdots \\
0 & 0 & 0 & \del_1^V & \cdots & 0 & 0 & \id_{V_0} & \cdots \\
\vdots & \vdots & \vdots & \vdots & \ddots & \vdots & \vdots & \vdots & \ddots \\
\end{pmatrix}.\]

We can then make the composition of two pushouts,
\[\begin{tikzcd}&V_\bullet \arrow[r, hookrightarrow] \arrow[d, hookrightarrow]& C_\bullet \arrow[d]\\
V_\bullet \arrow[d, hookrightarrow]\arrow[r, hookrightarrow] & W_\bullet \arrow[r] & R_\bullet \arrow[d]\arrow[ul, phantom, "\usebox\pushout", very near start]\\
D_\bullet \arrow[rr] & & T_\bullet\arrow[ul, phantom, "\usebox\pushout", very near start]\end{tikzcd}\]

The diagram is drawn with two different instances of $V_\bullet$ so that the diagram commutes.

The first inclusion of $V_\bullet$ into $W_\bullet$ sends $V_1$ into the 1st copy of $V_1$ in $W_1$ and the same for $V_0$ in $W_0$. The second inclusion of $V_\bullet$ sends $V_1$ into the $(r+1)$th copy of $V_1$ in $W_1$ and the same for $V_0$ in $W_0$. The inclusions of $V_\bullet$ into $C_\bullet$ and $D_\bullet$ are just inherited from the monic span above. As all these inclusions are basis-preserving, the code $T_\bullet$ can be uniquely defined up to relabelling of basis elements \cite[Lemma~5.4]{CowBu}, so all weight-related notions such as code distance, being $\omega$-limited etc. are canonical.

In this way we make the merged code $T_\bullet$ from the initial codes $C_\bullet$ and $D_\bullet$, where two logical operators in $C_\bullet$ and $D_\bullet$ respectively have been quotiented into the same equivalence class, performing a $\overline{Z}\otimes \overline{Z}$ measurement. This is done purely by initialising new qubits and stabilisers, so can be done in a fully error-corrected fashion, assuming the merge retains the code distance.

This is not generally guaranteed, so we must check it separately. It is also possible to introduce new logical qubits when doing this merge, for reasons described in \cite[Sec. C]{Coh}. These new logicals are often of low weight, so it can be useful to switch to a subsystem code \cite{KLP}, labelling the newly introduced logicals as gauge qubits. As we shall demonstrate, this frequently lets us increase the minimum distance of the merged code, which is now the minimum dressed distance of the subsystem code. When the depth $r = d$, the minimum distance of the codes beforehand, we assert that an external merge always maintains the code distance, when viewed as a subsystem code. For brevity we do not prove this here, but claim that it can be done by converting to the Tanner graph formalism and using similar arguments as in \cite[Sec. IV]{Coh} pertaining to `cleaning' \cite{BT}.

Increasing the depth $r$ of the code $P_\bullet$ will increase the size of $W_\bullet$ and hence the number of new data qubits and stabilisers added to the code. We would like to do this if a low depth results in a low distance. 

\subsubsection{Single-qubit measurements}\label{sec:single_qubit}

In \cite{Coh} new tensor product codes are also adjoined to the initial codes to perform logical single-qubit measurements. We will now convert this protocol into the homological picture. They also use their framework to perform logical multi-qubit Pauli measurements. We omit these as they are harder to view in the homological picture, although for those measurements which still yield CSS codes we assert that it can be done.

For single-qubit measurements, we only need one pushout. Again, say we are performing a $\overline{Z}$ measurement. Given an irreducible logical operator, we will make a new tensor product code.

\begin{definition}
Let $S_\bullet = \F_2^r \rightarrow \F_2^r$ be the classical code with parity-check matrix
\[\del_1^S = \begin{pmatrix}
1 & 0 & 0 & \cdots & 0 & 0\\
1 & 1 & 0 & \cdots & 0 & 0\\
0 & 1 & 1 & \cdots & 0 & 0\\
\vdots & \vdots & \vdots & \ddots & \vdots & \vdots \\
0 & 0 & 0 & \cdots & 1 & 1
\end{pmatrix}\]
with 1s on diagonal elements, and 1s on the entries below the diagonal, apart from the bottom-right diagonal entry which has no entry below it. 
\end{definition}

This is the incidence matrix of a `truncated' path graph, where the last vertex has been removed but its dangling incident edge remains. For example, if $r = 3$ then the graph is
\[\input{truncated_path_graph.tikz}\]
with 
\[\del_1^S = \begin{pmatrix}
1 & 0 & 0 \\
1 & 1 & 0 \\
0 & 1 & 1 \\
\end{pmatrix}\]

Observe that $\ker(\del_1^S) = \ker((\del_1^S)^\intercal = 0$, i.e. the classical code and its dual have no codespace, for any $r$.

Given an irreducible logical operator subcomplex for a CSS code $C_\bullet$, we can then make the code $(V \otimes S)_\bullet$. Similar to with external merges above, we have an inclusion $V_\bullet \hookrightarrow (V \otimes S)_\bullet$, where $V_1$ is sent to the first copy of $V_1$ in $(V \otimes S)_1$, and $V_0$ is sent to the first copy of $V_0$ in $(V \otimes S)_0$.

We then have a basis-preserving monic span
\[\begin{tikzcd}
V_\bullet \arrow[r, hookrightarrow, "g_\bullet"]\arrow[d, hookrightarrow, "f_\bullet"'] & (V \otimes S)_\bullet\\
C_\bullet & &
\end{tikzcd}\]
and so we can construct a new code by a single pushout
\[\begin{tikzcd}V_\bullet \arrow[r, hookrightarrow] \arrow[d, hookrightarrow]& (V \otimes S)_\bullet \arrow[d]\\
C_\bullet \arrow[r] & R_\bullet \arrow[ul, phantom, "\usebox\pushout", very near start]
\end{tikzcd}.\]

This time, $R_\bullet$ is the final code we are left with. We have initialised new qubits and stabilisers as dictated by $(V \otimes S)_\bullet$. As $(V \otimes S)_\bullet$ has no logical qubits, by the K{\"u}nneth formula, we have quotiented the $\overline{Z}$ logical operator $v$, which was used to construct $V_\bullet$ and so $(V \otimes S)_\bullet$, into the $[0]$ equivalence class. In other words, measuring the stabilisers of the new code will also perform a $\overline{Z}$ measurement on that logical qubit. As before, it is possible to incidentally introduce new logical qubits in the process.

If $r = d$, the distance of the initial code $C_\bullet$, then $R_\bullet$ will always have minimum dressed distance $d$ when viewed as a subsystem code \cite[Thm. 1]{Coh}, setting new logical qubits to be gauge qubits. If $r < d$ then this can still be the case, but it is not guaranteed. We will show in later sections that it is common to be able to perform such logical single-qubit measurements without requiring high depth $r$.

\subsubsection{Internal merges}

We can also perform surgery within a single codeblock $C_\bullet$, taking two logical $\overline{Z}$ operators from different logical qubits and merging them together. As before, we start with an irreducible logical operator subcomplex. This time, however, we have the diagram
\[\begin{tikzcd}V_\bullet \arrow[r, hookrightarrow,"f_\bullet" above, shift left=1.5ex]\arrow[r, hookrightarrow,"g_\bullet" below, shift right=1.5ex] & C_\bullet \end{tikzcd}\]
where $f_\bullet$ and $g_\bullet$ are basis preserving, and $\im(f_\bullet) \cap \im(g_\bullet) = 0$, i.e. there are no data qubits in the two logical operators which overlap, and the same for $X$ stabilisers.
\begin{remark}
It is possible to relax this condition of no overlap in certain situations, but we do not have a general theory of how to relax it, so we adhere to the condition here.
\end{remark}

We can then construct the merged code using the same tensor product code $W_\bullet = (P\otimes V)_\bullet$ from Section~\ref{sec:ext_merges}. This time, the merged code is the result of two coequalisers:
\[\begin{tikzcd}
V_\bullet \arrow[r, hookrightarrow, shift left=1.5ex]\arrow[r, hookrightarrow, shift right=1.5ex] & (W \oplus C)_\bullet \arrow[r] & R_\bullet \arrow[r] & T_\bullet\\
& V_\bullet \arrow[ur, hookrightarrow, shift left=1.5ex]\arrow[ur, hookrightarrow, shift right=1.5ex] & &
\end{tikzcd}\]
As with external merges, the diagram is drawn with two separate instances of $V_\bullet$ so that the diagram commutes.

The first two inclusions on the left take $V_\bullet$ and map it into $W_\bullet$ and $C_\bullet$ respectively. The $C_\bullet$ inclusion is $f_\bullet$, and the $W_\bullet$ inclusion takes $V_1$ and $V_0$ to their first copies in $W_1$ and $W_0$, as with internal merges.

The second two inclusions of $V_\bullet$ into $R_\bullet$ are as follows. One is $g_\bullet$ composed with the inclusion $C_\bullet \rightarrow R_\bullet$, and the second is the $W_\bullet$ inclusion taking $V_1$ and $V_0$ to their $(r+1)$th copies in $W_1$ and $W_0$, composed with the inclusion $W_\bullet \rightarrow R_\bullet$.

The intuition is we glue first one side of $W_\bullet$ into $C_\bullet$ based on the irreducible logical operator, then the same thing with the other side. It may be instructive to instead view the two coequalisers as a single pushout as follows:

\[\begin{tikzcd}(V\oplus V)_\bullet \arrow[r, hookrightarrow] \arrow[d, hookrightarrow]& W_\bullet \arrow[d]\\
C_\bullet \arrow[r] & T_\bullet \arrow[ul, phantom, "\usebox\pushout", very near start]
\end{tikzcd}\]

where the same data is contained in the universal construction. The inclusion $(V\oplus V)_\bullet \hookrightarrow W_\bullet$ takes one $V_\bullet$ to the first copy in $W_\bullet$, and the second $V_\bullet$ to the $(r+1)$th copy. The inclusion $(V\oplus V)_\bullet \hookrightarrow C_\bullet$ maps each $V_\bullet$ to the chosen logical operators to merge.

In \verb|SSIP|, the merged code is constructed using the two coequalisers diagram, so we stick with this picture.

Of course, we can view any external merge as an internal merge by setting $C_\bullet = (D \oplus E)_\bullet$ for some pair of codeblocks $D_\bullet$, $E_\bullet$. As for external merges, when the depth $r = d$, the minimum distance of the codes beforehand, we assert that an internal merge always maintains the code distance, when viewed as a subsystem code.

\section{Automated external surgery}\label{sec:auto_external}

We can now explain how \verb|SSIP| applies these universal constructions to perform surgery and extract useful data from merged codes. We start with external surgery, present results for external surgeries, then move on to internal surgery.

The basic data given to Algorithm~\ref{alg:ext_merge} for performing external surgery is as follows:
\begin{itemize}
\item The parity-check matrices of the two codes $C_\bullet$, $D_\bullet$ to be merged.
\item The two irreducible logicals $u \in C_1$ and $v\in D_1$ we would like to merge.
\item The basis ($Z$ or $X$) to perform the merge in.
\item The desired depth $r$ of the merge.
\end{itemize}

The codes are entered as \verb|CSScode| objects, while the logicals are vectors, and the depth is an unsigned integer. Verifying that a vector is an irreducible logical is straightforward linear algebra and is efficient to calculate, so we do not include this in the algorithm. We assume that the chosen basis is $Z$; as always, the $X$ version can be obtained by dualising to cochain complexes.

\begin{algorithm}
\caption{External merge calculation}\label{alg:ext_merge}
\begin{algorithmic}
\State $RM_1 \gets {\rm RestrictedMatrix}(u, \del_1^C)$
\State $RM_2 \gets {\rm RestrictedMatrix}(v, \del_1^D)$
\State ${\rm Span} \gets {\rm FindMonicSpan}(RM_1, RM_2)$
\If {Span is None}
  \State \Return None
\EndIf
\State $V_\bullet \gets RM_1$
\State $P_\bullet \gets {\rm ConstructP}(r)$
\State $W_\bullet \gets (P\otimes V)_\bullet$
\State ${\rm NewSpan1} \gets {\rm LHSspan(Span}, W_\bullet)$
\State ${\rm NewSpan2} \gets {\rm RHSspan(Span}, W_\bullet)$
\State $R_\bullet \gets {\rm Pushout}(V_\bullet, W_\bullet, C_\bullet, {\rm NewSpan1})$
\State $T_\bullet \gets {\rm Pushout}(V_\bullet, R_\bullet, D_\bullet, {\rm NewSpan2})$
\State \Return $T_\bullet$
\end{algorithmic}
\end{algorithm}

Let us explain this algorithm in more detail. \verb|RestrictedMatrix| simply takes a vector $u$ in $C_1$ and the differential $\del_1: C_1 \rightarrow C_0$ and calculates $R_1 = \del_1 \restriction_{{\rm supp\ } u}$, by removing columns with no support in $u$, and then removing any all-zero rows.

\verb|FindMonicSpan| is more interesting. There exists a basis-preserving monic span 
\[\begin{tikzcd}
V_\bullet \arrow[r, hookrightarrow, "f_\bullet"]\arrow[d, hookrightarrow, "g_\bullet"'] & C_\bullet\\
D_\bullet & &
\end{tikzcd}\]
if (but not only if) there are permutation matrices $M$, $N$ such that $R_1 = MR_2N$. That is, we have two injections $U_\bullet \hookrightarrow C_\bullet$ and $V_\bullet \hookrightarrow D_\bullet$, and we wish to find a basis-preserving isomorphism $U_\bullet \cong V_\bullet$ such that
we have an injection $V_\bullet \cong U_\bullet \hookrightarrow C_\bullet$. The basis-preserving isomorphism is given precisely by the permutation matrices $M$ and $N$, which dictate where basis elements of $V_1$ and $V_0$ are mapped to. The isomorphism explicitly is
\[\begin{tikzcd}V_1 \arrow[r]\arrow[d, "\sim"] & V_0 \arrow[d, "\sim"]\\
U_1 \arrow[r] & U_0\end{tikzcd}\]

Finding such permutation matrices is the hypergraph isomorphism problem. This in turn can be expressed as a graph isomorphism problem between bipartite graphs \cite{ADK}, at the cost of some increased space. The graph isomorphism problem is neither known to be poly-time nor NP-complete, but in practice is very fast to solve using VF2 \cite{CFSV}. \verb|SSIP| uses NetworkX \cite{HSS} to represent the graphs and call VF2.

\begin{remark}
We do not need a hypergraph isomorphism, only a hypergraph inclusion, to construct a basis-preserving monic span. However, isomorphism is necessary for $V_\bullet$ to be a logical operator subcomplex, see Definition~\ref{def:log_op_subcomplex}, in both codes. We rely on this property to perform logical parity measurements so throughout we assume our monic spans are constructed by hypergraph isomorphisms, and hence the two logicals to be merged are identical up to relabelling of qubits and checks.
\end{remark}

There may not be a graph isomorphism, in which case we do not find a basis-preserving monic span. In this case Algorithm~\ref{alg:ext_merge} returns \verb|None|. On the other hand, there may be many graph isomorphisms. In this case Algorithm~\ref{alg:ext_merge} just uses the first one found for simplicity. At times we may know the monic span \textit{a priori}, in which case this step can be skipped.

Once the monic span has been found the algorithm then performs the two pushouts. Taking a pushout of a basis-preserving monic span with a logical operator subcomplex at the apex is straightforward. Take the first pushout:
\[\begin{tikzcd}V_\bullet \arrow[r, hookrightarrow] \arrow[d, hookrightarrow]& C_\bullet \arrow[d]\\
W_\bullet \arrow[r] & R_\bullet \arrow[ul, phantom, "\usebox\pushout", very near start]\end{tikzcd}\]
We can expand this into components:
\[\begin{tikzcd}
V_0 \arrow[rrrrr, "f_0"]\arrow[ddddd, "l_0"] & & & & & C_0 \arrow[ddddd]\\
&V_1\arrow[rrr,"f_1"]\arrow[ddd, "l_1"]\arrow[ul, "\del^{V}_1"] & & & C_1\arrow[ddd] \arrow[ur, "\del^{C}_{1}"] &\\
& & 0\arrow[ul, "0"]\arrow[r, "0"]\arrow[d, "0"'] & C_2\arrow[ur, "\del^{C}_{2}"]\arrow[d] & &\\
& & W_{2}\arrow[dl,"\del^{W}_{2}"]\arrow[r] & R_{2}\arrow[dr, dotted, "\del^{R}_{2}"]\arrow[ul, phantom, "\usebox\pushout", very near start] & &\\
& W_1\arrow[rrr]\arrow[dl, "\del^{W}_1"] & & & R_1 \arrow[dr, dotted, "\del^{R}_{1}"]& \\
W_0 \arrow[rrrrr] & & & & & R_0
\end{tikzcd}\]
The pushout at degree 2 is just $R_2 = W_2 \oplus C_2$. At degree 1 we have $R_1 = W_1 \oplus C_1/ \im(l_1) \sim \im(f_1)$. To construct $\del_2^{R}$ we therefore start with $\del_2^{W}\oplus \del_2^{C}$ and add the rows corresponding to quotiented basis elements in $R_1$ together. That is, if $e_i$ is an entry in $V_1$, take the entries $l_1(e_i)$ and $f_1(e_i)$ and add those rows together in $\del_2^{W}\oplus \del_2^{C}$. All rows to be added together have disjoint support, so the addition of rows is unambiguous.

At degree 0 we have $R_0 = W_0 \oplus C_0/ \im(l_0) \sim \im(f_0)$. To construct $\del_1^R$ start with $\del_1^{W}\oplus \del_1^{C}$ then add rows corresponding to quotiented basis elements in $R_0$ together. Then, take the bitwise OR (logical inclusive) of columns corresponding to quotiented basis elements in $R_1$ together.

We verify in Appendix~\ref{app:comp_pushouts} that these differentials are the mediating maps given by the universal properties of pushouts, and that the above diagram commutes. The second pushout to acquire $T_\bullet$ follows in the same fashion.

Optionally, Algorithm~\ref{alg:ext_merge} can calculate some additional data to inform the user what the effect of the merge has been. This is wrapped up into a \verb|MergeResult| object. In addition to the output \verb|CSScode|, this object contains:
\begin{itemize}
\item The inclusion matrix $C_1 \oplus D_1 \hookrightarrow T_1$, i.e. the map on qubits from the initial codes to the merged code.
\item The row indices for any new $Z$ stabilisers initialised in $P_Z$.
\item The row indices for any new $X$ stabilisers initialised in $P_X$.
\item The indices of any new qubits initialised.
\item A basis for any new $\overline{Z}$ logical operators introduced.
\item A basis for any new $\overline{X}$ logical operators introduced.
\item A basis for the $k_C + k_D - 1$ $\overline{Z}$ logical operators inherited from $C_\bullet$ and $D_\bullet$, which we call `old' $\overline{Z}$ logical operators, as opposed to the `new' $\overline{Z}$ logical operators which can be incidentally introduced during a merge.
\item A basis for the `old' $\overline{X}$ logical operators.
\end{itemize}

We know the inclusion matrix immediately from the pushouts; the same is true for the indices of new stabilisers and qubits. Calculating the new and old logical operators is done by calculating logicals in the initial codes and multiplying through by the inclusion matrix, then taking the appropriate quotient to find the new logicals.

Overall, aside from the graph isomorphism problem all of the subroutines in this section have at worst $\CO(n^3)$ runtime, with the worst complexity being Gaussian elimination, which is required for the additional data. We find in practice that graph isomorphism is not a bottleneck using VF2. For codes with hundreds of qubits Algorithm~\ref{alg:ext_merge} runs in a few seconds or at most minutes on a Mac laptop. Given that most of the time is spent doing linear algebra in Python, should the runtime become problematic then implementation in a faster language such as C should let Algorithm~\ref{alg:ext_merge} run in seconds for codes with many thousands of qubits, until the graph isomorphism problem becomes challenging.

In practice, there is some hidden complexity here. Given two arbitrary CSS codes with no additional knowledge of the code structure, the problem to solve is not just whether, given two irreducible logical operators, we can perform an external merge. We would have to work out which irreducible logical operators are available, and so could be paired up to merge. Assuming we have no additional knowledge this will be a formidable problem in general: the number of logical operators will typically scale exponentially with the blocklength of the codes, and the number of possible pairings of logicals between the codes will explode combinatorially. Even if running Algorithm~\ref{alg:ext_merge} is extremely fast, the combinatorial explosion makes exhaustively finding all monic spans between two large codes implausible.

Thus for codes of high blocklength we would like to know in advance the structure of the available irreducible logical operators. Fortunately, modern qLDPC codes are not random, and in fact tend to be highly structured, such as lifted product codes. In \cite{BCGMRY, ES} this structure is leveraged to find irreducible $\overline{Z}$ and $\overline{X}$ logicals for every qubit.

\subsection{Small examples}

We warm up to our results on external surgery with some small $d=3$ codes with $k=1$ each. We will use combinations of the
\begin{itemize}
\item $\llbracket 9, 1, 3\rrbracket$ Shor code \cite{Shor},
\item $\llbracket 15, 1, 3\rrbracket$ Quantum Reed-Muller (QRM) code \cite{KLZ},
\item $\llbracket 7, 1, 3\rrbracket$ Steane code \cite{St},
\item $\llbracket 9, 1, 3 \rrbracket$ rotated surface code \cite{BM2}, and
\item $\llbracket 13, 1, 3\rrbracket$ unrotated surface code \cite{Kit},
\end{itemize}
which we call our small example set.

In addition to all having distance 3, these codes have the following property: there exists a weight 3 $\overline{Z}$ logical operator $v$ with the restricted matrix
\[\del^{C}_{1}\restriction_{{\rm supp\ } v}\hspace{2mm} \sim \begin{pmatrix}
1 & 1 & 0\\
0 & 1 & 1\\
\end{pmatrix}\]
where $\sim$ means up to permutation of rows and columns. The restricted matrix is the parity-check matrix of a repetition code; this will always be true for an irreducible logical operator, as the only non-zero element in $\ker(\del^{C_\bullet}_{1}\restriction_{{\rm supp\ } v})$ must be the all-1s vector, and every minimum-weight logical operator must be irreducible.

The above restricted matrix has $d-1$ rows (once all-zero rows have been removed). We will always have a monic span with $V_\bullet$ having the differential above. Thus we can always do external surgery between any two of these codes, and the merged codes with $r=1$ will have 2 additional data qubits when compared to the disjoint initial codes, i.e. $\dim T_1 = \dim C_1 + \dim D_1 + 2$. The same applies for $\delta^1_C \restriction_{{\rm supp\ } v}$ for $\overline{X}$ logicals instead, with the exception of the QRM code which has $d_X = 7$ so has no weight 3 $\overline{X}$ logicals.

Of course, not every restricted matrix of a weight $d$ logical has $d-1$ rows for any other $d=3$ CSS code, as there may be redundant checks on that logical. An example for which Algorithm~\ref{alg:ext_merge} would fail to find any monic spans for $\overline{Z}$ logicals with the codes in our small example set, despite having distance 3, is the $\llbracket 18, 2, 3 \rrbracket$ toric code. The restricted matrix for a weight 3 $\overline{Z}$ logical in the toric code is
\[\begin{pmatrix}
1 & 1 & 0\\
0 & 1 & 1\\
1 & 0 & 1
\end{pmatrix}\]
up to permutation, i.e. there is an extra $X$-check; the dual applies for a weight 3 $X$ logical, which will have an extra $Z$-check.

As it turns out, we can do surgery between any two of the codes in our example set in the $Z$ basis with depth $r=1$ while maintaining $d=3$ in the merged codes. All of the merged codes have 1 logical qubit. For the $X$ basis the same applies with the exception of the QRM code.

The only remaining figure of merit is $\omega$, the maximum weight of any column or row. We show in Figure~\ref{fig:small_merges} that we increase $\omega$ by at most 1 when compared to the codes beforehand. We do not claim that this is optimal -- we can obviously do $X$ merges between distance 3 surface codes without increasing $\omega$, but this will depend on the choice of logical used.

\begin{figure}
\begin{center}
\begin{tabular}{|c||c|c|c|c|c|}
\hline
$\omega_{\rm after} - \omega_{\rm before}$ & Shor & QRM & Steane & Rotated surface & Surface \\
\hhline{|=||=|=|=|=|=|}
Shor & 1, 0 & 1 & 1, 0 & 1, 0 & 1, 0 \\
\hline
QRM & & 1 & 1 & 1 & 1 \\
\hline
Steane & & & 1, 1 & 1, 1 & 1, 1  \\
\hline
Rotated surface & & & & 1, 1 & 1, 1  \\
\hline
Surface & & & & & 0, 1  \\
\hline
\end{tabular}
\end{center}
\caption{$\omega_{\rm after} - \omega_{\rm before}$ for merges between weight 3 logicals. Values for $Z$ merges are shown first and values for $X$ merges second, when weight 3 logicals exist.}\label{fig:small_merges}
\end{figure}

Interestingly, the QRM code is triorthogonal \cite{BHaah}, meaning that it admits a transversal logical $T$ gate. This means that one can use \verb|SSIP| to generate merges between a triorthogonal code and some other code to inject $T$ states. One candidate for the other code is the surface code, with which one can easily perform Cliffords \cite{BLKW}. The QRM code is a small example of a 3D colour code \cite{Bom}, and the merging protocol scales to arbitrary distance $d$ 3D colour codes and surface codes, using only $d-1$ additional qubits to perform the injection. 3D colour codes and surface codes are perhaps not the best candidates for fault-tolerant computation for reasons of threshold and code parameter scaling compared to other qLDPC codes, but the principle is interesting, and different from Bombin's code-switching protocol \cite{Bom}.

It is unsurprising that one can do surgery with our small example set. They can all be seen as topological codes. Other than those already mentioned, the Shor code is a tessellation of $\R P^2$ \cite{FM} and the Steane code is a 2D colour code \cite{BM3}. In fact, surgery between 2D colour codes and surface codes has already been described in \cite{NFB}.

We could also perform logical single-qubit measurements with our small example set using the method in Section~\ref{sec:single_qubit}, but this is uninteresting as every code in our set has $k=1$ so measuring the logical qubit in the $Z$ or $X$ basis can be done by measuring out every data qubit in the $Z$ or $X$ basis. Similarly, it does not make sense to do internal surgery with $k=1$ codeblocks.

\subsection{Lift-connected surface codes}\label{sec:lcs_codes}

Lift-connected surface (LCS) codes \cite{ORM} are lifted product codes where the commutative matrix subring is $\mathscr{C}_\ell$, the ring of $\ell$-by-$\ell$ circulant matrices. Intuitively, one can think of LCS codes as disjoint surface codes which are then connected by some stabilisers. They are interesting in part because their parameters can outperform those of surface codes even at low blocklengths. They also perform comparably to surface codes against phenomenological noise, and can be implemented with 3D local connectivity.

LCS codes are constructed using two variables: $\ell$, the size of the circulant matrices, and $L$, the length of the `base' code over $\mathscr{C}_\ell$.\footnote{In this paper we have swapped round the variable labelling compared to the original LCS paper \cite{ORM} in order to conform to the notation in \cite{PK3}.}

Let $P^{(0)} = \id_\ell$ and $P^{(1)}$ be the first right cyclic shift of $\id_\ell$, so for example
\[P^{(1)} = \begin{pmatrix}
0 & 1 & 0\\
0 & 0 & 1\\
1 & 0 & 0
\end{pmatrix}\]
when $\ell = 3$.

Then, construct the $L$-by-$L+1$ matrix
\[B = \begin{pmatrix}
P^{(0)} & P^{(0)} + P^{(1)} & 0 & \cdots & 0 & 0\\
0 & P^{(0)} & P^{(0)} + P^{(1)} & \cdots & 0 & 0\\
\vdots & & & \ddots & \vdots & \vdots \\
0 & 0 & \cdots & 0 & P^{(0)} & P^{(0)} + P^{(1)} 
\end{pmatrix}\]
and let $A = B^\intercal$. Take the tensor product over $\mathscr{C}_\ell$, recalling from Section~\ref{sec:tensor_recap} that this gives
\[P_Z = \begin{pmatrix}B \underset{\mathscr{C}_\ell}{\otimes} \id_{L+1} &\id_{L} \underset{\mathscr{C}_\ell}{\otimes} A\end{pmatrix}; \quad P_X= \begin{pmatrix} \id_{L+1} \underset{\mathscr{C}_\ell}{\otimes} B & A \underset{\mathscr{C}_\ell}{\otimes} \id_{L}\end{pmatrix}\]
which we then view over $\F_2$.

LCS codes have parameters $\llbracket ((L+1)^2 + L^2)\ell, \ell, \min(\ell, 2L+1) \rrbracket$. Strictly speaking, the parameter $d =\min(\ell, 2L+1)$ has not been proven, just conjectured with empirical evidence to support it. Under this conjecture, LCS codes have distance scaling linearly in $n$ until $\ell = 2L+1$. This is shown to be true for LCS codes under a certain size \cite[Sec.~III A]{ORM}. LCS codes are also $\omega$-limited with $\omega = 6$, so they are qLDPC codes.

\subsubsection{Individual merges}\label{sec:lcs_ind_merges}

We test individual merges between LCS codes, without considering parallelisation. We take the set of LCS codes with $L \in \{1,2,3\}$ and $\ell \in \{L+2, L+3, L+4\}$, except we truncate at $\ell=6$. Our smallest initial code has $L = 1$, $\ell = 3$ and parameters $\llbracket 15, 3, 3\rrbracket$. Our largest initial code has $L = 3$, $\ell = 6$ and parameters $\llbracket 150, 6, 6 \rrbracket$. These blocklengths are chosen such that our results are reproducible in a few hours on a personal computer, and so that the initial codes are close to having the best possible parameters for LCS codes, see \cite[Fig.~4]{ORM}.

Our method is simple. For each code $C_\bullet$, we find an arbitrary tensor product decomposition of the logical space, i.e. a basis of the homology space $H_1(C_\bullet)$, and its consistent basis for the cohomology space $H^1(C^\bullet)$. We then pick out representative logicals for each one, which we call ${u_i}$ for $u_i \in [u_i]$, recalling that the basis set is $\{[u_i]\}_{u \in I}$, such that the logicals are irreducible. In principle, should we not find an irreducible logical for a given qubit we would leave that qubit out, but we successfully find irreducible logicals for all qubits in our benchmarking set.

We then test by taking two identical copies of $C_\bullet$ and merging them along their shared irreducible logical $u_i$. Evidently this is guaranteed to give a monic span as the two codes are identical. After the merge, there may be additional logical qubits introduced. We relegate these to being gauge qubits and make the merged code a subsystem CSS code.

For every merge, we compute three figures of merit: (1) the depth $r$ required for the merges to leave the code distance unchanged, so $d = \min(\ell, 2L+1)$ when viewed as a subsystem code, (2) the total number of additional data qubits required as a proportion of the total length of the original codes $n_{\rm ancilla}/{n_{\rm initial}}$, (3) the maximum weight of any row or column in the parity-check matrices $\omega$. In Figure~\ref{fig:Z_ind_merges_lcs} we present the mean average of these values over each of the $\overline{X}$ and $\overline{Z}$ logicals for a given pair of codes. For example, the $\llbracket 15, 3, 3\rrbracket$ code has 3 logical qubits, so we perform 3 different $X$-merges and average out the values of $r$, $n_{\rm ancilla}/{n_{\rm initial}}$ and $\omega$ for the 3 different merged codes. For more fine-grained results, where we show the results of each merge rather than just their averages, see Appendix~\ref{app:more_results}. The scripts for running all benchmarks can be found at \verb|https://github.com/CQCL/SSIP/benchmarks|. We explicitly calculated all subsystem code distances using Z3, as the distance is low enough for this method to be practical.

\begin{figure}
\begin{center}
\begin{tabular}{|c|c||c|c|c|}
\hline
$L$ & $\ell$ & $\langle r\rangle$ & $\langle n_{\rm ancilla}/{n_{\rm initial}} \rangle$ & $\langle \omega \rangle$ \\
\hhline{|=|=||=|=|=|}
1 & 3 & 1 & $0.16$ & 6\\
\hline
1 & 4 & 1 & $0.14$ & 6 \\
\hline
1 & 5 & 1 & $0.12$ & 6 \\
\hline
2 & 4 & 1 & $0.1$ & 7 \\
\hline
2 & 5 & 2 & $0.36$ & 7 \\
\hline
2 & 6 & 2 & $0.3$ & 7 \\
\hline
3 & 5 & 2.6 & $0.37$ & 7 \\
\hline
3 & 6 & 2.8 & $0.31$ & 7 \\
\hline
\end{tabular}

\vspace{5mm}

\begin{tabular}{|c|c||c|c|c|}
\hline
$L$ & $\ell$ & $\langle r\rangle$ & $\langle n_{\rm ancilla}/{n_{\rm initial}} \rangle$ & $\langle \omega \rangle$ \\
\hhline{|=|=||=|=|=|}
1 & 3 & 1 & $0.15$ & 6\\
\hline
1 & 4 & 1 & $0.125$ & 6 \\
\hline
1 & 5 & 1 & $0.112$ & 6 \\
\hline
2 & 4 & 1.75 & $0.25$ & 7 \\
\hline
2 & 5 & 1.4 & $0.16$ & 7 \\
\hline
2 & 6 & 1.6 & $0.14$ & 7 \\
\hline
3 & 5 & 1.4 & $0.14$ & 7 \\
\hline
3 & 6 & 1.7 & $0.15$ & 7 \\
\hline
\end{tabular}
\end{center}
\caption{Figures of merit for individual $X$ and $Z$ merges between LCS codes.}\label{fig:Z_ind_merges_lcs}
\end{figure}

Reading through the first table of Figure~\ref{fig:Z_ind_merges_lcs}, we can see first that the average required depth $r$ increases as the size of the initial codes increases. This is not surprising, as the larger the initial code the more likely it is that performing a low depth merge with another code will incidentally add a logical, which does not belong wholly to the new logical qubits introduced, with a lower distance. Regardless, the depth required remains low, with the maximum required for any of the merges being 4. Similarly, as the size of the initial codes increases so does the typical proportion of new qubits required for the merges. The intermediate code being added is a tensor product code, which itself has vanishing rate and poorer distance scaling when compared to the underlying quantum memories. Thus when we have to increase the depth $r$ we are adding more of a `worse' code, slightly inhibiting the efficiency of the merges.

In the second table of Figure~\ref{fig:Z_ind_merges_lcs} the depth does not increase significantly, nor does the ancilla qubit overhead. We do not have a convincing structural explanation for this discrepancy between $Z$s and $X$s. It appears that our rudimentary technique for finding logicals happens to find particularly suitable ones in the $Z$-merge case, and less suitable ones in the $X$-merge case as the blocklengths increase. This highlights the sensitivity of the surgeries to the logicals used.

There are cases at larger sizes where the merges can be performed at $r=1$, and in these cases $n_{\rm ancilla}/{n_{\rm initial}}$ is extremely low. For instance, for $L=3, \ell = 5$, which makes $n_C = 125$, so $n_{\rm initial} = 250$, there is a $\overline{Z}$ logical which gives an $r=1$ merge at $n_{\rm ancilla}/{n_{\rm initial}} = 0.048$, a marginal overhead of 12 ancillary data qubits compared to the overall blocksize of 250. We expect that a greater understanding of the structure of LCS codes would yield more logicals which admit low depth merges, but our rudimentary technique only finds these occasionally.

Lastly, $\omega$ remains virtually constant throughout, at just above $\omega = 6$ for the initial LCS codes. In \cite[Lem.~5.18]{CowBu} we showed that merges of qLDPC codes remain qLDPC, and for LCS codes the row and column weights barely increase.

In Figure~\ref{fig:Z_LCS_compare} we also compare to both lattice surgery and a naive application of \cite{Coh} to performing the same merges in the $Z$ basis. That is, for lattice surgery we generate $2k$ (unrotated) surface code patches with the same $d$ as the LCS codes, then perform a single merge between two patches and record the total overhead. For \cite{Coh} we use LCS codes, but then for merging we initialise large $r=d$ tensor product ancillae codes which are connected appropriately.

As mentioned in Section~\ref{ref:related}, comparisons between these different procedures will generally be apples-to-oranges. The results here are for generating codeblocks and then performing a single merge between them. The advantage of surface codes is that one can easily parallelise lattice surgery, while the same is not true of our protocol. For example, constructing 12 surface code patches, as in the $L = 3$, $\ell =6$ comparison case, and then performing a single merge between two of them is quite a contrived scenario. Similarly, for the approach of \cite{Coh} one can do Pauli measurements using any combination of logical qubits, while ours is more restricted. We also used the default procedure, where the ancilla blocks of \cite{Coh} have high depth, where in reality one could perhaps get away with reducing the overhead while maintaining distance.

With those caveats out of the way, the main take away of Figure~\ref{fig:Z_LCS_compare} is that for individual merges at low blocklength, our homological approach requires much less overhead than the other two methods. A common feature of both our procedure and that of \cite{Coh} is that, while the initial quantum memories have better parameters than those of surface codes, some of the overhead is instead offloaded onto the ancillae used for surgery. Surface codes require very small numbers of additional qubits.

\begin{figure}
\begin{center}
\begin{tabular}{|c|c||c|c|c|}
\hline
$L$ & $\ell$ & $n_{\rm initial}$ & $\langle n_{\rm ancilla} \rangle$ & $\langle n_{\rm total} \rangle$ \\
\hhline{|=|=||=|=|=|}
1 & 3 & 30 & 4.5 & 34.5 \\
  &   & 78 & 2   & 80   \\
  &   & 30 & 49  & 79   \\
\hline
1 & 4 & 40 & 5.6 & 45.6 \\
  &   & 104& 2   & 106  \\
  &   & 40 & 55  & 95  \\
\hline
1 & 5 & 50 & 5.6 & 55.6 \\
  &   & 130& 2   & 132  \\
  &   & 50 & 60.6& 110.6 \\
\hline
2 & 4 & 104 & 26  & 130 \\
  &   & 200 & 3   & 203 \\
  &   & 104 & 142 & 246 \\
\hline
2 & 5 & 130 & 20.8& 150.8\\
  &   & 410 & 4   & 414  \\
  &   & 130 & 199 & 329  \\
\hline
2 & 6 & 156 & 21.8& 177.8 \\
  &   & 492 & 4   & 496  \\
  &   & 156 & 209 & 365  \\
\hline
3 & 5 & 250 & 35  & 285 \\
  &   & 410 & 4   & 414  \\
  &   & 250 & 316.6 & 566.6  \\
\hline
3 & 6 & 300 & 45  & 345 \\
  &   & 732 & 5   & 737  \\
  &   & 300 & 404.3 & 704.3  \\
\hline
\end{tabular}
\end{center}
\caption{Comparison of LCS code individual $Z$-merges to surface codes and \cite{Coh}. The first row in each box is our homological approach using Algorithm~\ref{alg:ext_merge}. The second is lattice surgery with surface code patches. The third is a naive application of \cite{Coh} to LCS codes.}\label{fig:Z_LCS_compare}
\end{figure}

\subsubsection{Parallel merges}\label{sec:lcs_par_merges}

Now we present results for performing a logical merge between \textit{every} pair of logical qubits in the two codes simultaneously, as shown in the schematic below:
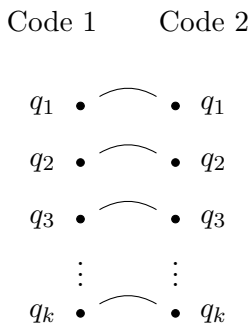
\[\input{merge_schema.tikz}\] 
We use the same logicals as before. This time, our approach is as follows: take $r=1$ for a merge between logical qubits of the same index. Should this result in a merged code with lower $d$ than the initial codes, when viewed as a subsystem CSS code, then increment $r$ to 2 for \textit{every} merge, and so on until $d = \min(\ell, 2L+1)$. This is to avoid having to explore the search space of different possible depths for each merge.

We follow the same procedure of using Z3 to calculate subsystem code distances. Results are presented in Figure~\ref{fig:Z_par_merges_lcs}.

\begin{figure}
\begin{center}
\begin{tabular}{|c|c||c|c|c|}
\hline
$L$ & $\ell$ & $r $ & $ n_{\rm ancilla}/{n_{\rm initial}} $ & $\omega $ \\
\hhline{|=|=||=|=|=|}
1 & 3 & 1 & $0.47$ & 8\\
\hline
1 & 4 & 1 & $0.55$ & 9 \\
\hline
1 & 5 & 1 & $0.62$ & 10 \\
\hline
2 & 4 & 2 & $1.27$ & 10 \\
\hline
2 & 5 & 1 & $0.52$ & 11 \\
\hline
2 & 6 & 1 & $0.54$ & 12 \\
\hline
3 & 5 & 2 & $1.34$ & 11 \\
\hline
3 & 6 & 2 & $1.37$ & 12 \\
\hline
\end{tabular}

\vspace{5mm}

\begin{tabular}{|c|c||c|c|c|}
\hline
$L$ & $\ell$ & $r $ & $ n_{\rm ancilla}/{n_{\rm initial}} $ & $\omega $ \\
\hhline{|=|=||=|=|=|}
1 & 3 & 1 & $0.43$ & 8\\
\hline
1 & 4 & 1 & $0.5$ & 9 \\
\hline
1 & 5 & 1 & $0.56$ & 10 \\
\hline
2 & 4 & 2 & $1.15$ & 9 \\
\hline
2 & 5 & 3 & $2.08$ & 10 \\
\hline
2 & 6 & 2 & $1.33$ & 11 \\
\hline
3 & 5 & 1 & $0.34$ & 11 \\
\hline
3 & 6 & 2 & $1.1$ & 11 \\
\hline
\end{tabular}
\end{center}
\caption{Figures of merit for parallel $X$ and $Z$ merges between LCS codes.}\label{fig:Z_par_merges_lcs}
\end{figure}

Here we see the cost of parallelisation. Not only does the overhead in terms of ancillae data qubits increase, so too does $\omega$. This is more-or-less unavoidable with efficient codes: the logicals being used to perform merges are likely to have overlap on some data qubits and stabilisers, so the new tensor product codes will increase stabiliser weights and the number of stabilisers some data qubits are in the support of. Despite this, the depths are encouraging. The largest LCS codes used have 6 logical qubits, but despite this only a merge depth of $r \leq 2$ was required for their parallel merges. We again compare the overhead required to that of surface codes with the same $k$ and $d$ as the LCS codes, and a naive application of Cohen et al. \cite{Coh} in Figure~\ref{fig:Z_LCS_par_compare}.

\begin{figure}
\begin{center}
\begin{tabular}{|c|c||c|c|c|}
\hline
$L$ & $\ell$ & $n_{\rm initial}$ & $ n_{\rm ancilla} $ & $ n_{\rm total} $ \\
\hhline{|=|=||=|=|=|}
1 & 3 & 30 & 13 & 43 \\
  &   & 78 & 6  & 84   \\
  &   & 30 & 147 & 177  \\
\hline
1 & 4 & 40 & 20 & 60 \\
  &   & 104& 6   & 110  \\
  &   & 40 & 220  & 260  \\
\hline
1 & 5 & 50 & 28 & 78 \\
  &   & 130& 6   & 136  \\
  &   & 50 & 303 & 353 \\
\hline
2 & 4 & 104 & 120  & 224 \\
  &   & 200 & 12   & 212 \\
  &   & 104 & 568 & 672 \\
\hline
2 & 5 & 130 & 271 & 401\\
  &   & 410 & 20   & 430  \\
  &   & 130 & 995 & 1125  \\
\hline
2 & 6 & 156 & 207& 363 \\
  &   & 492 & 24   & 416  \\
  &   & 156 & 1254 & 1410  \\
\hline
3 & 5 & 250 & 91  & 341 \\
  &   & 410 & 20   & 430  \\
  &   & 250 & 1583 & 1833  \\
\hline
3 & 6 & 300 & 331  & 631 \\
  &   & 732 & 30   & 762  \\
  &   & 300 & 2426 & 2726  \\
\hline
\end{tabular}
\end{center}
\caption{Comparison of LCS code parallel $Z$-merges to surface codes and \cite{Coh}. The first row in each box is our homological approach using Algorithm~\ref{alg:ext_merge}. The second is lattice surgery with surface code patches. The third is a naive application of \cite{Coh} to LCS codes.}\label{fig:Z_LCS_par_compare}
\end{figure}

While still comparing favourably in terms of overall qubit overhead to surface codes, the advantage is significantly weakened. This is because surface codes make it extremely easy to parallelise merges. We expect substantial gains could be made by considering the logicals used more carefully, and lowering the level of parallelisation somewhat without restricting ourselves to individual merges.

Naive application of \cite{Coh} performs very poorly by contrast. This is because the quantum memory is not yet large enough for its efficiency as a qLDPC code to outweigh the large ancilla requirements when compared to surface codes, and because we can `get away with' having low depth merges in our homological approach.

\subsubsection{Individual single-qubit measurements}\label{sec:lcs_ind_singleqs}

We now retrace our steps for the same benchmarking set but performing single-qubit logical measurements instead, following Section~\ref{sec:single_qubit}. Recall that we are claiming no novelty in our approach here, it is merely that of \cite{Coh} translated into chain complexes. It is still interesting, however, because in \cite[Table.~1]{Coh} the results given are just estimates at high depth. We show that it is possible to perform these single-qubit measurements while incurring lower overhead.

We show figures of merit for single-qubit measurements in Figure~\ref{fig:Z_ind_singleqs_lcs}. Overall, on the LCS benchmarking set they tend to be more expensive than individual external merges, both as a fraction of the initial blocklength and the raw number of ancilla qubits. Similarly, in Figure~\ref{fig:Z_LCS_ind_singleqs_compare} we see that while low depth measurements in this manner still compare favourably to those of surface codes, the difference is much less pronounced, and again one should bear in mind that surface codes favour parallelisation much better.

\begin{figure}
\begin{center}
\begin{tabular}{|c|c||c|c|c|}
\hline
$L$ & $\ell$ & $\langle r\rangle$ & $\langle n_{\rm ancilla}/{n_{\rm initial}} \rangle$ & $\langle \omega \rangle$ \\
\hhline{|=|=||=|=|=|}
1 & 3 & 1.67 & $0.8$ & 6\\
\hline
1 & 4 & 1.75 & $0.75$ & 6 \\
\hline
1 & 5 & 1.8 & $0.7$ & 6 \\
\hline
2 & 4 & 1.5 & $0.5$ & 7 \\
\hline
2 & 5 & 3 & $1.18$ & 7 \\
\hline
2 & 6 & 2.5 & $0.78$ & 7 \\
\hline
3 & 5 & 3.2 & $0.84$ & 7 \\
\hline
3 & 6 & 3.5 & $0.97$ & 7 \\
\hline
\end{tabular}

\vspace{5mm}

\begin{tabular}{|c|c||c|c|c|}
\hline
$L$ & $\ell$ & $\langle r\rangle$ & $\langle n_{\rm ancilla}/{n_{\rm initial}} \rangle$ & $\langle \omega \rangle$ \\
\hhline{|=|=||=|=|=|}
1 & 3 & 1.67 & $0.7$ & 6\\
\hline
1 & 4 & 1.75 & $0.75$ & 6 \\
\hline
1 & 5 & 1.8 & $0.64$ & 6 \\
\hline
2 & 4 & 1.2 & $0.3$ & 7 \\
\hline
2 & 5 & 2.2 & $0.58$ & 7 \\
\hline
2 & 6 & 2.17 & $0.51$ & 7 \\
\hline
3 & 5 & 2.2 & $0.52$ & 7 \\
\hline
3 & 6 & 2.67 & $0.58$ & 7 \\
\hline
\end{tabular}
\end{center}
\caption{Figures of merit for individual single-qubit logical $X$ and $Z$ measurements with LCS codes.}\label{fig:Z_ind_singleqs_lcs}
\end{figure}

\begin{figure}
\begin{center}
\begin{tabular}{|c|c||c|c|c|}
\hline
$L$ & $\ell$ & $n_{\rm initial}$ & $\langle n_{\rm ancilla} \rangle$ & $\langle n_{\rm total} \rangle$ \\
\hhline{|=|=||=|=|=|}
1 & 3 & 15 & 10.5 & 25.5 \\
  &   & 39 & 0  & 39   \\
  &   & 15 & 23 & 38  \\
\hline
1 & 4 & 20 & 15 & 35 \\
  &   & 52 & 0   & 52  \\
  &   & 20 & 26  & 46  \\
\hline
1 & 5 & 25 & 16 & 41 \\
  &   & 65 & 0   & 65 \\
  &   & 25 & 28.8 & 53.8 \\
\hline
2 & 4 & 52 & 15.6  & 77.6 \\
  &   & 100 & 0   & 100 \\
  &   & 52 & 69 & 121 \\
\hline
2 & 5 & 65 & 37.7 & 102.7\\
  &   & 205 & 0   & 205  \\
  &   & 65 & 97 & 162  \\
\hline
2 & 6 & 78 & 39.8 & 117.8 \\
  &   & 246 & 0   & 246  \\
  &   & 78 & 102 & 180  \\
\hline
3 & 5 & 125 & 65  & 190 \\
  &   & 205 & 0   & 205  \\
  &   & 125 & 155.8 & 280.8  \\
\hline
3 & 6 & 150 & 87  & 237 \\
  &   & 366 & 0   & 366  \\
  &   & 150 & 199 & 349  \\
\hline
\end{tabular}
\end{center}
\caption{Comparison of LCS code individual single-qubit logical $Z$-measurements to surface codes and a naive application of \cite{Coh}. The first row uses the method described in Section~\ref{sec:single_qubit}. The second is lattice surgery with surface code patches. The third is a naive application of \cite{Coh} to LCS codes.}\label{fig:Z_LCS_ind_singleqs_compare}
\end{figure}

\subsubsection{Parallel single-qubit measurements}\label{sec:lcs_par_singleqs}

We could consider performing single-qubit measurements in the same basis on \textit{every} logical qubit in parallel, but this would be a completely contrived benchmark, as this can always be done in a CSS code by measuring out the existing data qubits, rather than adding new data qubits. Instead, we consider the following: take the first half (rounded down) of the logical qubits and perform logical single-qubit measurements on these in parallel. The half of the logical qubits is chosen arbitrarily. We show figures of merit for doing this in Figure~\ref{fig:par_singleqs_lcs}, then show comparisons to surface codes and a naive application of \cite{Coh} in Figure~\ref{fig:Z_LCS_par_singleqs_compare}.

\begin{figure}
\begin{center}
\begin{tabular}{|c|c||c|c|c|}
\hline
$L$ & $\ell$ & $r$ & $ n_{\rm ancilla}/{n_{\rm initial}} $ & $ \omega$ \\
\hhline{|=|=||=|=|=|}
1 & 3 & 2 & 1.0 & 6\\
\hline
1 & 4 & 2 & 1.65 & 7\\
\hline
1 & 5 & 2 & 1.56 & 7\\
\hline
2 & 4 & 2 & 1.37 & 8\\
\hline
2 & 5 & 3 & 1.6 & 8\\
\hline
2 & 6 & 3 & 2.24 & 9\\
\hline
3 & 5 & 2 & 1.03 & 8\\
\hline
3 & 6 & 2 & 1.33 & 9\\
\hline
\end{tabular}

\vspace{5mm}

\begin{tabular}{|c|c||c|c|c|}
\hline
$L$ & $\ell$ & $r$ & $ n_{\rm ancilla}/{n_{\rm initial}}$ & $\omega $ \\
\hhline{|=|=||=|=|=|}
1 & 3 & 2 & 1.07 & 6\\
\hline
1 & 4 & 2 & 1.85 & 7\\
\hline
1 & 5 & 2 & 1.6 & 7\\
\hline
2 & 4 & 3 & 2.0 & 8\\
\hline
2 & 5 & 2 & 0.98 & 8\\
\hline
2 & 6 & 2 & 1.27 & 9\\
\hline
3 & 5 & 2 & 0.58 & 8\\
\hline
3 & 6 & 3 & 1.49 & 9\\
\hline
\end{tabular}
\end{center}
\caption{Figures of merit for parallel single-qubit logical $X$ and $Z$ measurements with LCS codes.}\label{fig:par_singleqs_lcs}
\end{figure}

\begin{figure}
\begin{center}
\begin{tabular}{|c|c||c|c|c|}
\hline
$L$ & $\ell$ & $n_{\rm initial}$ & $ n_{\rm ancilla} $ & $n_{\rm total}$ \\
\hhline{|=|=||=|=|=|}
1 & 3 & 15 & 16 & 31 \\
  &   & 39 & 0  & 39   \\
  &   & 15 & 27 & 42  \\
\hline
1 & 4 & 20 & 37 & 57 \\
  &   & 52 & 0   & 52  \\
  &   & 20 & 62  & 82  \\
\hline
1 & 5 & 25 & 40 & 65 \\
  &   & 65 & 0   & 65 \\
  &   & 25 & 67 & 92 \\
\hline
2 & 4 & 52 & 104 & 156 \\
  &   & 100 & 0   & 100 \\
  &   & 52 & 145 & 197 \\
\hline
2 & 5 & 65 & 64 & 129\\
  &   & 205 & 0   & 205  \\
  &   & 65 & 190 & 255  \\
\hline
2 & 6 & 78 & 99 & 177 \\
  &   & 246 & 0   & 246  \\
  &   & 78 & 294 & 372  \\
\hline
3 & 5 & 125 & 73  & 198 \\
  &   & 205 & 0   & 205  \\
  &   & 125 & 217 & 342  \\
\hline
3 & 6 & 150 & 224  & 374 \\
  &   & 366 & 0   & 366  \\
  &   & 150 & 491 & 641  \\
\hline
\end{tabular}
\end{center}
\caption{Comparison of LCS code parallel single-qubit logical $Z$-measurements to surface codes and a naive application of \cite{Coh}. The first row uses the method described in Section~\ref{sec:single_qubit}. The second is lattice surgery with surface code patches. The third is a naive application of \cite{Coh} to LCS codes.}\label{fig:Z_LCS_par_singleqs_compare}
\end{figure}

In Appendix~\ref{app:GB_codes} we rerun this entire benchmarking procedure for generalised bicycle (GB) codes \cite{PK3,KoPr}. We find a similar story there, but GB codes are even more amenable to surgery and compare extremely favourably compared to surface codes and the approach of \cite{Coh}.

\subsection{The gross code}\label{sec:ext_gross}

Bivariate bicycle (BB) codes \cite{BCGMRY} are lifted products over the ring $\mathscr{C}_{\ell}\otimes \mathscr{C}_m$, that is the tensor product over rings of circulant matrices of different dimensions, but viewed over $\F_2$.

Recalling that $\mathscr{C}_{\ell} \cong \F_2^{\langle \ell \rangle}$, it is immediate that 
\[\mathscr{C}_{\ell} \otimes \mathscr{C}_m \cong \F_2^{\langle \ell \rangle} \otimes \F_2^{\langle m \rangle},\]
the ring of polynomials over two variables $x$ and $y$ modulo $x^\ell -1$ and $y^m - 1$. The bijection sends 
\[P^{(1)}_{\ell} \otimes \id_m \mapsto x;\quad \id_{\ell} \otimes P^{(1)}_m \mapsto y,\]
and $x^\ell = y^m = 1$. Recall that $P^{(1)}_{\ell}$ is the right cyclic shift matrix of $\id_{\ell}$, as in Section~\ref{sec:lcs_codes}.

Then, let
\[A = A_1 + A_2 + A_3; \quad B = B_1 + B_2 + B_3\]
where each matrix $A_i$ and $B_j$ is a power of $x$ or $y$, interpreted in $\mathscr{C}_{\ell} \otimes \mathscr{C}_m$. Bivariate bicycle (BB) codes have $P_X = \begin{pmatrix}A & B\end{pmatrix}$ and $P_Z = \begin{pmatrix}B^\intercal & A^\intercal \end{pmatrix}$ for some matrices $A$ and $B$. Therefore each BB code is uniquely defined by a pair of polynomials in $\F_2^{\langle \ell \rangle} \otimes \F_2^{\langle m \rangle}$, each of which is the sum of three monomials in a single variable. Observe that $n = 2\ell m$ for any BB code. Call the first and second block of $\ell m$ data qubits the `unprimed' and `primed' blocks respectively.

The example we focus on in this work is the ``gross code", a $\llbracket 144, 12 ,12 \rrbracket$ code with $\ell = 12$, $m = 6$, $A = x^3 + y + y^2$ and $B = y^3 + x + x^2$. This code has a high error threshold under circuit-level noise, its Tanner graph can be decomposed into two planar subgraphs, which is important for planar architectures, and it is $\omega$-limited with $\omega = 6$.

As we are only using one code here, the benchmarking we perform will be a bit more exhaustive. For this, we use another useful feature of BB codes: we can calculate a basis of the logical space using the algebraic structure of the codes. In the case of the gross code this gives us an immediate set of logical Paulis with weight 12, one $\overline{Z}$ and one $\overline{X}$ for each logical qubit. As $d=12$, these logicals are also irreducible. We forgo further details but see \cite[Sec.~9.1]{BCGMRY}. We use these logicals for all our benchmarking in this section. These logicals split into primed and unprimed sets, with the (un)primed set having support only in the (un)primed block.

First we find that there is a basis-preserving monic span between every pair of $\overline{Z}$ logicals in the primed block; the same applies to every pair of logicals in the unprimed block, and also to $\overline{X}$ logicals. Therefore, given two copies of the gross code we can perform individual external merges between any of the logical qubits which belong to the same block, in either basis. As $d=12$ for the gross code, checking preservation of distance in merged codes is out of reach of the Z3 algorithm in reasonable compute time, so we again rely on \verb|QDistRnd|. The upshot is that these individual merges can each be done with a depth $r=1$, requiring only 18 additional data qubits and increasing $\omega$ to 7, leaving the code distance as a subsystem code at 12 assuming the bound from \verb|QDistRnd| is tight.

Furthermore, we find that we can do parallel merges on all 12 logical qubits between two copies of the gross code in either basis using only $r=1$, requiring a total of $18\times 12 = 216$ extra data qubits. This raises $\omega$ to 12.

Similarly we can study individual single-qubit logical measurements. We find that single-qubit logical $X$ measurements on the unprimed block require a depth of $r=3$, and so 78 extra data qubits. They also introduce 72 new stabiliser generators, so a total of 150 new qubits including syndrome qubits. $X$ measurements on the primed block require only a depth of $r=1$, 18 extra data qubits, and 12 new generators so a total of 30 new qubits. These all raise $\omega$ to 7. These overheads are far below that required by performing such measurements naively, as it was predicted in \cite[Sec.~9.4]{BCGMRY} that these measurements would each require a total of 1380 extra qubits when including syndrome qubits (although the authors did expect this value to be optimised significantly). The flipped version applies to single-qubit logical $Z$ measurements: those in the unprimed block require a depth of $r=1$, and a total of 30 new qubits. Those in the primed block require $r=3$ and 150 total new qubits.

Additionally, we can perform parallel single-qubit logical measurements. We can measure every logical qubit in the unprimed block in the $X$ basis with $r=3$ using 468 new data qubits and 432 new syndrome qubits, so 900 ancillae in total. This increases $\omega$ to 8. The same applies to the primed block in the $Z$ basis.

We can measure every logical qubit in the primed block in the $X$ basis with $r=1$ using 108 new data qubits and 72 new syndrome qubits, so 190 ancillae in total. This increases $\omega$ to 11. The same applies to the unprimed block in the $Z$ basis.

\section{Automated internal surgery}\label{sec:auto_internal}
As mentioned earlier we can also use \verb|SSIP| to perform internal surgery, that is surgery between logicals in the same codeblock. This is performed in a similar manner to external surgery but with some minor differences. The basic data given to Algorithm~\ref{alg:int_merge} is as follows:
\begin{itemize}
\item The parity-check matrices of $C_\bullet$, the code within which internal surgery will be performed.
\item The two irreducible logicals $u, v \in C_1$ we would like to merge.
\item The basis ($Z$ or $X$) to perform the merge in.
\item The desired depth $r$ of the merge.
\end{itemize}

\begin{algorithm}
\caption{Internal merge calculation}\label{alg:int_merge}
\begin{algorithmic}
\State $RM_1 \gets {\rm RestrictedMatrix}(u, \del_1^C)$
\State $RM_2 \gets {\rm RestrictedMatrix}(v, \del_1^C)$
\If {$|RM_1 \cap RM_2| \neq 0$}
  \State \Return None
\EndIf
\State ${\rm Diagram} \gets {\rm FindDiagram}(RM_1, RM_2)$
\If {Diagram is None}
  \State \Return None
\EndIf
\State $V_\bullet \gets RM_1$
\State $P_\bullet \gets {\rm ConstructP}(r)$
\State $W_\bullet \gets (P\otimes V)_\bullet$
\State ${\rm NewDiagram1} \gets {\rm LHSdiagram(Diagram}, W_\bullet)$
\State $R_\bullet \gets {\rm Coequaliser}(V_\bullet, (W \oplus C)_\bullet, {\rm NewDiagram1})$
\State ${\rm NewDiagram2} \gets {\rm RHSdiagram(Diagram}, W_\bullet)$
\State $T_\bullet \gets {\rm Coequaliser}(V_\bullet, R_\bullet, {\rm NewDiagram2})$
\State \Return $T_\bullet$
\end{algorithmic}
\end{algorithm}

The first thing Algorithm~\ref{alg:int_merge} does is calculate the restrictions of $P_X$ to the support of the logicals $u$ and $v$. It then finds $|RM_1 \cap RM_2|$, by which we mean the set of data qubits which have overlapping support, and the same for stabiliser generators. If there is any overlap on either of these, Algorithm~\ref{alg:int_merge} rejects the merge and outputs \verb|None|. The algorithm then proceeds similarly to Algorithm~\ref{alg:ext_merge}: it computes a hypergraph isomorphism between the restricted matrices, then computes a tensor product code to merge the two logicals together. Finally, it computes the two coequalisers and returns the merged code $T_\bullet$. See Appendix~\ref{app:comp_pushouts} for this computation. Optionally, Algorithm~\ref{alg:int_merge} can return a \verb|MergeResult| object, which contains the same additional data as in Algorithm~\ref{alg:ext_merge}.

\subsection{The gross code}

We return to the $\llbracket 144, 12, 12 \rrbracket$ gross code to conduct benchmarking on internal merges. While we are guaranteed to have a diagram of the form
\[\begin{tikzcd}V_\bullet \arrow[r, hookrightarrow,"f_\bullet" above, shift left=1.5ex]\arrow[r, hookrightarrow,"g_\bullet" below, shift right=1.5ex] & C_\bullet \end{tikzcd}\]
whenever $u$ and $v$ are in the set of irreducible logicals given in \cite[Sec.~9.1]{BCGMRY} and belong to the same block (primed or unprimed), these will commonly have some overlap in data qubits or stabilisers. Therefore we cannot perform internal merges using any arbitrary pair of logical qubits in the same block, as we could with external merges in Section~\ref{sec:ext_gross}.

\begin{remark}
Of course, each logical qubit has many irreducible logicals associated to it, and so we could try to find pairs which do not overlap. This is a large search space so we just stick with the irreducible logicals given in \cite[Sec.~9.1]{BCGMRY}. 
\end{remark}

In Figure~\ref{fig:int_merges_gross} we show results for internal merges in the $X$ and $Z$ basis. All of the possible merges increase $\omega$ to 7. At depth $r=1$ we use 18 ancilla data qubits for a merge. For $r=2$ we use 48, and for $r=3$ we use 78.

Of the possible 15 different internal merges one could do within a primed or unprimed block, we find that 12 different $X$ merges can be done in the unprimed block, as the logicals have no overlap, while none can be done in the primed block. For $Z$ merges, only 3 can be done in the unprimed block, while 12 can be done in the primed block.

\begin{figure}
\begin{center}
\begin{tabular}{|c||c|c|c|c|c|c|c|c|c|c|c|c|}
\hline
$r$ & 0 & 1 & 2 & 3 & 4 & 5 & 6 & 7 & 8 & 9 & 10 & 11 \\
\hhline{|=||=|=|=|=|=|=|=|=|=|=|=|=|}
0   & - & 2 & 2 & 2 & 3 & - & - & - & - & - & - & - \\
\hline
1   & 2 & - & - & 2 & 3 & 2 & - & - & - & - & - & - \\
\hline
2   & 2 & - & - & 3 & 2 & 2 & - & - & - & - & - & - \\
\hline
3   & 2 & 2 & 3 & - & - & 2 & - & - & - & - & - & - \\
\hline
4   & 3 & 3 & 2 & - & - & 2 & - & - & - & - & - & - \\
\hline
5   & - & 2 & 2 & 2 & 2 & - & - & - & - & - & - & - \\
\hline
6   & - & - & - & - & - & - & - & - & - & - & - & - \\
\hline
7   & - & - & - & - & - & - & - & - & - & - & - & - \\
\hline
8   & - & - & - & - & - & - & - & - & - & - & - & - \\
\hline
9   & - & - & - & - & - & - & - & - & - & - & - & - \\
\hline
10  & - & - & - & - & - & - & - & - & - & - & - & - \\
\hline
11  & - & - & - & - & - & - & - & - & - & - & - & - \\
\hline
\end{tabular}

\vspace{5mm}

\begin{tabular}{|c||c|c|c|c|c|c|c|c|c|c|c|c|}
\hline
$r$ & 0 & 1 & 2 & 3 & 4 & 5 & 6 & 7 & 8 & 9 & 10 & 11 \\
\hhline{|=||=|=|=|=|=|=|=|=|=|=|=|=|}
0   & - & - & - & - & - & 1 & - & - & - & - & - & - \\
\hline
1   & - & - & - & - & - & 1 & - & - & - & - & - & - \\
\hline
2   & - & - & - & - & - & - & - & - & - & - & - & - \\
\hline
3   & - & - & - & - & - & 1 & - & - & - & - & - & - \\
\hline
4   & - & - & - & - & - & - & - & - & - & - & - & - \\
\hline
5   & 1 & 1 & - & 1 & - & - & - & - & - & - & - & - \\
\hline
6   & - & - & - & - & - & - & - & 3 & - & 2 & 2 & 2 \\
\hline
7   & - & - & - & - & - & - & 3 & - & 3 & 2 & 2 & - \\
\hline
8   & - & - & - & - & - & - & - & 3 & - & 2 & 3 & 2 \\
\hline
9   & - & - & - & - & - & - & 2 & 2 & 2 & - & - & 2 \\
\hline
10  & - & - & - & - & - & - & 2 & 2 & 3 & - & - & 3 \\
\hline
11  & - & - & - & - & - & - & 2 & - & 2 & 2 & 3 & - \\
\hline
\end{tabular}

\end{center}
\caption{Depths required for individual internal $X$ and $Z$ merges between logical qubits $i$ and $j$ in the gross code. Dashed entries have no internal merge for the logical operators chosen.}\label{fig:int_merges_gross}
\end{figure}

\section{Future directions}

In order for surgeries identified with \verb|SSIP| to be useful in practice we must tackle the problems which \verb|SSIP| does not handle, as stated in the introduction, namely: establishing (pseudo-)thresholds for codes throughout the surgery process, along with circuits for the syndrome measurements, and decoders which function throughout. 

There are recently developed classes of lifted product qLDPC codes which we have not tested \verb|SSIP| on, for which it could be interesting to do so \cite{LP,SHR}. Beyond these, it would be very interesting to consider the design of efficient qLDPC CSS codes which, in addition to other useful properties such as low depth syndrome circuits, also admit surgery between and within codeblocks with low overhead of additional qubits, while provably retaining high code distance. This would allow us to avoid the problems of (a) trying to find suitable logicals to construct merges, which results in a combinatorial explosion when done naively, and (b) calculating distance after merges, which is always going to be difficult without additional \textit{a priori} knowledge of the code's structure.

In a different direction, while the chain complex formalism is perhaps more sophisticated than \textit{ad hoc} constructions with topological codes, the actual tensor product codes we are using to perform merges are quite primitive; they are the obvious generalisation of the small codes used to merge patches in lattice surgery. There is no reason why there should not be more sophisticated ancilla codes which could be initialised to merge logicals, going beyond just parity measurements to many-qubit measurements in certain codes, but which do not suffer from the higher overhead of ancillae used for many-qubit measurements in \cite{Coh}.

The most interesting use-case for surgery is merging different classes of codes, such that we can achieve universality using the different logical operations available natively to the codes. To that end it is an interesting question to consider large codes which are triorthogonal or otherwise admit transversal logical non-Clifford gates, and which have low thresholds and favourable code parameters. Given such large codes, performing code merges could allow us to cheaply teleport magic states into quantum memories which admit Clifford operations, or vice versa, thereby circumventing Eastin-Knill \cite{EK} without requiring magic-state distillation. Previous resource estimates using code-switching protocols with the 3D colour code have indicated that distillation is superior \cite{BeKuSv}, but code-switching is somewhat different to magic state injection by surgery, and those limitations may not apply when using different triorthogonal codes.

\section{Acknowledgements}
We thank Seyon Sivarajah for his valuable input on the codebase and comments on the paper. We are grateful to Simon Burton for contributing the code to lower bound the minimum distance using Z3, and for his initial conception of constructing new CSS codes by taking pushouts. We also thank the Wolfson Harrison UK Research Council Quantum Foundation Scholarship for making this work possible.

\input{appendix}
\bibliographystyle{eptcs}
\end{document}

%% file: basic.tikzstyles
\tikzstyle{boundary vertex}=[inner sep=0mm, minimum size=1mm, shape=circle, draw=black, fill=black]
\tikzstyle{grey_dot}=[fill={rgb,255: red,191; green,191; blue,191}, draw={rgb,255: red,191; green,191; blue,191}, shape=circle, minimum size=1mm, inner sep=0mm]
\tikzstyle{blue_dot}=[fill={rgb,255: red,202; green,251; blue,255}, draw=black, shape=circle, minimum size=1mm, inner sep=0mm]
\tikzstyle{white_dot}=[fill=white, draw=black, shape=circle, minimum size=1.5mm, inner sep=0mm]

\tikzstyle{arrow}=[->]
\tikzstyle{red_arrow}=[->, draw=red]
\tikzstyle{cyan_arrow}=[->, draw=cyan]
\tikzstyle{red_dash}=[-, dashed, draw=red]
\tikzstyle{grey dash}=[-, fill=none, draw={rgb,255: red,191; green,191; blue,191}, dashed]
\tikzstyle{dashed arrow}=[->, dashed]
\tikzstyle{blue_edge}=[-, draw={rgb,255: red,46; green,126; blue,255}]
\tikzstyle{red_edge}=[-, draw=red]

%% file: quantum.tikzstyles
\tikzstyle{gate}=[shape=rectangle, text height=1.5ex, text depth=0.25ex, yshift=0.5mm, fill=white, draw=black, minimum height=5mm, yshift=-0.5mm, minimum width=5mm, font={\small}, tikzit category=circuit]
\tikzstyle{big gate}=[shape=rectangle, text height=1.5ex, text depth=0.25ex, yshift=0.5mm, fill=white, draw=black, minimum height=18mm, yshift=-0.5mm, minimum width=5mm, font={\small}, tikzit category=circuit]
\tikzstyle{Z dot}=[inner sep=0mm, minimum size=2mm, shape=circle, draw=black, fill={rgb,255: red,221; green,255; blue,221}, tikzit category=zx]
\tikzstyle{Z phase dot}=[minimum size=5mm, font={\footnotesize\boldmath}, shape=rectangle, rounded corners=2mm, inner sep=0.2mm, outer sep=-2mm, scale=0.8, tikzit shape=circle, draw=black, fill={rgb,255: red,221; green,255; blue,221}, tikzit draw=blue, tikzit category=zx]
\tikzstyle{X dot}=[Z dot, shape=circle, draw=black, fill={rgb,255: red,255; green,136; blue,136}, tikzit category=zx]
\tikzstyle{X phase dot}=[Z phase dot, tikzit shape=circle, tikzit draw=blue, fill={rgb,255: red,255; green,136; blue,136}, font={\footnotesize\boldmath}, tikzit category=zx]
\tikzstyle{hadamard}=[fill=yellow, draw=black, shape=rectangle, inner sep=0.6mm, minimum height=1.5mm, minimum width=1.5mm, tikzit category=zx]
\tikzstyle{paulibox}=[fill={rgb,255: red,221; green,221; blue,255}, draw=black, shape=rectangle, inner sep=0.6mm, minimum height=5mm, minimum width=5mm, font={\footnotesize}, text height=1.5ex, text depth=0.25ex, tikzit category=zx]
\tikzstyle{vertex}=[inner sep=0mm, minimum size=1mm, shape=circle, draw=black, fill=black, tikzit category=misc]
\tikzstyle{vertex set}=[inner sep=0mm, minimum size=1mm, shape=circle, draw=black, fill=white, font={\footnotesize\boldmath}, tikzit category=misc]
\tikzstyle{small black dot}=[fill=black, draw=black, shape=circle, inner sep=0pt, minimum width=1.2mm, tikzit category=circuit]
\tikzstyle{cnot ctrl}=[fill=black, draw=black, shape=circle, inner sep=0pt, minimum width=1.2mm, tikzit category=circuit]
\tikzstyle{cnot targ}=[fill=white, draw=white, shape=circle, tikzit category=circuit, label={center:$\oplus$}, inner sep=0pt, minimum width=2.1mm, tikzit fill={rgb,255: red,102; green,204; blue,255}, tikzit draw=black]
\tikzstyle{ket}=[fill=white, draw=black, shape=regular polygon, regular polygon sides=3, regular polygon rotate=-30, scale=0.7, inner sep=1pt, tikzit category=circuit, tikzit shape=rectangle, tikzit fill=green]
\tikzstyle{bra}=[fill=white, draw=black, shape=regular polygon, regular polygon sides=3, regular polygon rotate=30, scale=0.7, inner sep=1pt, tikzit category=circuit, tikzit shape=rectangle, tikzit fill=red]
\tikzstyle{scalar}=[shape=rectangle, text height=1.5ex, text depth=0.25ex, yshift=0.5mm, fill=white, draw=black, minimum height=5mm, yshift=-0.5mm, minimum width=5mm, font={\small}]
\tikzstyle{clabel}=[fill=white, draw=none, shape=rectangle, tikzit fill={rgb,255: red,56; green,255; blue,242}, font={\footnotesize}, inner sep=1pt, tikzit category=labels]
\tikzstyle{empty diagram}=[draw={gray!40!white}, dashed, shape=rectangle, minimum width=1cm, minimum height=1cm, tikzit category=misc]

\tikzstyle{hadamard edge}=[-, dashed, dash pattern=on 2pt off 0.5pt, thick, draw={rgb,255: red,68; green,136; blue,255}]
\tikzstyle{box edge}=[-, dashed, dash pattern=on 2pt off 0.5pt, thick, draw={rgb,255: red,203; green,192; blue,225}]
\tikzstyle{brace edge}=[-, tikzit draw=blue, decorate, decoration={brace,amplitude=1mm,raise=-1mm}]
\tikzstyle{diredge}=[->]
\tikzstyle{double edge}=[-, double, shorten <=-1mm, shorten >=-1mm, double distance=2pt]
\tikzstyle{gray edge}=[-, {gray!60!white}]
\tikzstyle{pointer edge}=[->, very thick, gray]
\tikzstyle{boldedge}=[-, line width=1.6pt, shorten <=-0.17mm, shorten >=-0.17mm]

%% file: truncated_path_graph.tikz
\begin{tikzpicture}
	\begin{pgfonlayer}{nodelayer}
		\node [style=boundary vertex] (0) at (-3, 0) {};
		\node [style=boundary vertex] (1) at (-1, 0) {};
		\node [style=boundary vertex] (2) at (1, 0) {};
		\node [style=none] (3) at (3, 0) {};
	\end{pgfonlayer}
	\begin{pgfonlayer}{edgelayer}
		\draw (0) to (1);
		\draw (1) to (2);
		\draw (2) to (3.center);
	\end{pgfonlayer}
\end{tikzpicture}

%% file: merge_schema.tikz
\begin{tikzpicture}
	\begin{pgfonlayer}{nodelayer}
		\node [style=boundary vertex] (0) at (-1.25, 3) {};
		\node [style=boundary vertex] (1) at (-1.25, 1.5) {};
		\node [style=boundary vertex] (2) at (-1.25, 0) {};
		\node [style=boundary vertex] (3) at (-1.25, -2.5) {};
		\node [style=boundary vertex] (4) at (1.25, 3) {};
		\node [style=boundary vertex] (5) at (1.25, 1.5) {};
		\node [style=boundary vertex] (6) at (1.25, 0) {};
		\node [style=boundary vertex] (7) at (1.25, -2.5) {};
		\node [style=none] (8) at (-1.25, -1.25) {$\vdots$};
		\node [style=none] (9) at (1.25, -1.25) {$\vdots$};
		\node [style=none] (10) at (-0.75, 3.25) {};
		\node [style=none] (11) at (0.75, 3.25) {};
		\node [style=none] (12) at (-2.25, 3) {$q_1$};
		\node [style=none] (13) at (-2.25, 0) {$q_3$};
		\node [style=none] (14) at (-2.25, 1.5) {$q_2$};
		\node [style=none] (15) at (-2.25, -2.5) {$q_k$};
		\node [style=none] (16) at (2.25, 3) {$q_1$};
		\node [style=none] (17) at (2.25, 0) {$q_3$};
		\node [style=none] (18) at (2.25, 1.5) {$q_2$};
		\node [style=none] (19) at (2.25, -2.5) {$q_k$};
		\node [style=none] (20) at (-2, 5.25) {Code 1};
		\node [style=none] (21) at (2, 5.25) {Code 2};
		\node [style=none] (23) at (-0.75, 1.75) {};
		\node [style=none] (24) at (0.75, 1.75) {};
		\node [style=none] (25) at (-0.75, 0.25) {};
		\node [style=none] (26) at (0.75, 0.25) {};
		\node [style=none] (27) at (-0.75, -2.25) {};
		\node [style=none] (28) at (0.75, -2.25) {};
	\end{pgfonlayer}
	\begin{pgfonlayer}{edgelayer}
		\draw [bend left] (10.center) to (11.center);
		\draw [bend left] (23.center) to (24.center);
		\draw [bend left] (25.center) to (26.center);
		\draw [bend left] (27.center) to (28.center);
	\end{pgfonlayer}
\end{tikzpicture}

%% file: appendix.tex
\appendix

\section{Subsystem code distance calculation}\label{app:sub_distance}

Given a CSS code $C_\bullet$ defined by two parity-check matrices $P_X \in \F_2^{m_X\times n}$, $P_Z \in \F_2^{m_Z\times n}$ and a function $f$ which calculates or estimates the distance of a CSS code, we show that the same function $f$ can be called to calculate or estimate the distance of a subsystem CSS code. For simplicity we assume that $f$ can yield $d_Z$, the $Z$-distance, or $d_X$, the $X$-distance, as desired, as this is what \verb|DistRandCSS| can do \cite{QDR}. However, the method works even if $f$ outputs $d = \min(d_Z,d_X)$.

First, find a spanning set of the gauge $\overline{Z}$ logicals $\CG \subset H_1(C_\bullet)$. Typically this set will be a basis, but it does not have to be. Say the set chosen has cardinality $l_Z$. Then append these logicals to $P_Z$, making a new matrix $P'_Z \in \F_2^{(m_Z+l_Z)\times n}$. These gauge logicals must commute with the $X$ stabilisers, so we have a new CSS code $C'_\bullet$ with parity-check matrices $P_X$ and $P'_Z$. Apply $f$ to $C'_\bullet$ to acquire $d_Z'$. This is the lowest weight $\overline{Z}$ logical which is not in the image of $P'_Z$, and so is the lowest weight dressed $\overline{Z}$ logical in our subsystem code.

Then do the same the other way round. Make a new matrix $P'_X \in \F_2^{(m_X+l_X)\times n}$, giving a new CSS code $C''_\bullet$ with parity-check matrices $P'_X$ and $P_Z$. Apply $f$ to acquire $d'_X$. Then the distance of our subsystem CSS code is $d' = \min(d'_Z, d'_X)$.

\section{Computing colimits}\label{app:comp_pushouts}
Here we show explicitly how to calculate the colimits necessary for Algorithms~\ref{alg:ext_merge} and \ref{alg:int_merge}. We only need to check it for coequalisers, as an external merge can always be viewed as an internal merge within a direct sum codeblock.

We start off with the following diagram

\[\begin{tikzcd}
   & V_2 \arrow[r]\arrow[ddl, "g_2"' near start] & V_1\arrow[ddl, "g_1"' near start] \arrow[r] & V_0\arrow[ddl, "g_0"' near start] \\
   V_2 \arrow[r]\arrow[d, "f_2"'] & V_1 \arrow[r]\arrow[d, "f_1"'] & V_0\arrow[d, "f_0"'] & \\
   R_2 \arrow[r]\arrow[d, "{\rm coeq}_2"] & R_1 \arrow[r]\arrow[d, "{\rm coeq}_1"] & R_0\arrow[d, "{\rm coeq}_0"] & \\
   T_2 \arrow[r] & T_1 \arrow[r] & T_0 & \\
\end{tikzcd}\]

and our task is to find $\del_2^T$ and $\del_1^T$. We know from the universal property that once the components $T_i$ are fixed, the mediating maps are also unique, so we need only find differentials such that the diagram commutes.

Recall that $g_\bullet$ and $f_\bullet$ are basis-preserving chain maps, and ${\rm coeq}_i := {\rm coeq}(f_i, g_i)$ is the coequaliser of matrices at degree $i$, which is also basis-preserving. $V_\bullet$ is simultaneously the operator subcomplex of two irreducible $\overline{Z}$ logicals in $R_\bullet$. The first observation is that for all merges in the $Z$ basis, $V_2 = 0$, hence $f_2 = g_2 = 0$ and $T_2 = R_2$. Therefore $\del_2^T = {\rm coeq_1}\circ \del_2^R$. This can be computed simply be taking the XOR of pairs of rows in the mutual image of $g_1$ and $f_1$ and assigning the new rows entries in $\del_2^T$ -- it does not matter which so long as we are consistent.

For $\del_1^T$, we have that $\del_1^T\circ {\rm coeq_1} = {\rm coeq_0}\circ \del_1^R$. We can compute $\del_1^T\circ {\rm coeq_1}$ in the same way, by taking XORs of pairs of rows. We must now find $\del_1^T$. This time we take the OR (not XOR) of pairs of columns in $\del_1^T\circ {\rm coeq_1}$ which correspond to basis elements in the mutual image of $g_1$ and $f_1$. The assignment of new column entries must be consistent with the assignment of row entries to $\del_2^T$ so that ${\rm coeq_1}$ is identical to before.

One can check that the diagram now commutes everywhere, and so we have calculated the mediating maps, i.e. differentials of $T_\bullet$, successfully. When performing an external merge the matrices of $R_\bullet$ are block-diagonal and this procedure reduces to the computation described in Section~\ref{sec:auto_external}.

\section{Generalised bicycle codes}\label{app:GB_codes}

Generalised bicycle (GB) codes are codes with $P_X = \begin{pmatrix}A & B\end{pmatrix}$ and $P_Z = \begin{pmatrix}B^\intercal & A^\intercal \end{pmatrix}$ where $A$ and $B$ are elements of $\mathscr{C}_{\ell}$, the ring of circulant matrices \cite{PK3,KoPr}. GB codes are a special case of lifted product codes where the classical codes over $\mathscr{C}_{\ell}$ have the check matrices $A$ and $B$. In this sense they are smallest possible lifted product codes.

$\mathscr{C}_{\ell} \cong \F_2^{\langle \ell \rangle}$, so $A$ and $B$ can thus be described uniquely by polynomials over $\F_2$ in a single variable $x$, where $x^\ell = 1$.

There is no exact formula for all the parameters of GB codes, but certain input polynomials have been found to yield exceptional parameters. For example, $\ell = 63$ with $a(x) = 1+x+x^{14}+x^{16}+x^{22}$ and $b(x) = 1+x^3+x^{13}+x^{20}+x^{42}$ gives a $\llbracket 126, 28, 8\rrbracket$ code. In all our benchmarks here we use the GB codes (A1) to (A5) from \cite[App.~B]{PK3}, ranging from $n=46$ to $n = 254$. For codes (A1) and (A5) the distance is not known, with only upper and lower bounds given. We assume the upper bound is tight for the purposes of the paper. The distances are large enough that we do not check distances of merged codes using Z3 but instead rely solely on \verb|QDistRnd| to estimate.

\subsection{Individual merges}

We perform the same benchmarking as in Section~\ref{sec:lcs_ind_merges}, but with GB codes instead of LCS codes. The codes reach high dimensions, with up to 28 logical qubits, so we only perform merges using the first 7 logical qubits to prevent exorbitant compute time. As we are only using \verb|QDistRnd| to estimate the merged code distances we cannot prove that at the merge depths given the distance does not decrease, but this is somewhat justified as the distance is not even known for some of the initial codes.

\begin{figure}
\begin{center}
\begin{tabular}{|c||c|c|c|}
\hline
$\ell$ & $\langle r\rangle$ & $\langle n_{\rm ancilla}/{n_{\rm initial}} \rangle$ & $\langle \omega \rangle$ \\
\hhline{|=||=|=|=|}
23 & 1 & 0.23 & 9 \\
\hline
24 & 1 & 0.19 & 9 \\
\hline
63 & 1.29 & 0.35 & 11 \\
\hline
90 & 1.14 & 0.24 & 9 \\
\hline
127 & 1 & 0.24 & 11 \\
\hline
\end{tabular}

\vspace{5mm}

\begin{tabular}{|c||c|c|c|}
\hline
$\ell$ & $\langle r\rangle$ & $\langle n_{\rm ancilla}/{n_{\rm initial}} \rangle$ & $\langle \omega \rangle$ \\
\hhline{|=||=|=|=|}
23 & 1 & 0.22 & 9 \\
\hline
24 & 1.17 & 0.3 & 9 \\
\hline
63 & 1 & 0.25 & 11 \\
\hline
90 & 1 & 0.18 & 9 \\
\hline
127 & 1 & 0.23 & 11 \\
\hline
\end{tabular}
\end{center}
\caption{Figures of merit for individual $X$ and $Z$ merges between GB codes.}\label{fig:X_ind_merges_gb}
\end{figure}

Results are shown in Figure~\ref{fig:X_ind_merges_gb} for individual $X$ and $Z$ merges respectively. We also show a comparison with surface codes and \cite{Coh} in Figure~\ref{fig:Z_GB_compare}.

\begin{figure}
\begin{center}
\begin{tabular}{|c||c|c|c|}
\hline
$\ell$ & $n_{\rm initial}$ & $\langle n_{\rm ancilla} \rangle$ & $\langle n_{\rm total} \rangle$ \\
\hhline{|=||=|=|=|}
23 & 92 & 20.24 & 112.24 \\
   & 580 & 8   & 588   \\
   & 92 & 560  & 652   \\
\hline
24 & 96 & 28.8 & 124.8 \\
   & 1356 & 7  & 1363  \\
   & 96 & 541.33 & 637.33  \\
\hline
63 & 252 & 63 & 315 \\
  &  6328 & 7 & 6335  \\
  &  252 & 1526.86 & 1778.86 \\
\hline
90 & 360 & 64.8 & 424.8 \\
   & 12260 & 17   & 12277 \\
   & 360 & 3857.71 & 4217.71 \\
\hline
127 & 508 & 116.84 & 624.84 \\
    & 42616 & 19  & 42635  \\
    & 508 & 7493.71 & 8001.71  \\
\hline
\end{tabular}
\end{center}
\caption{Comparison of GB code individual $Z$-merges to surface codes and \cite{Coh}. The first row in each box is our homological approach using Algorithm~\ref{alg:ext_merge}. The second is lattice surgery with surface code patches. The third is a naive application of \cite{Coh} to GB codes.}\label{fig:Z_GB_compare}
\end{figure}

GB codes appear to be highly amenable to surgery, with extremely efficient individual merges compared to surface codes and \cite{Coh}. We cannot discount the possibility that \verb|QDistRnd| fails to find low weight logicals in the merged codes, however.

\subsection{Parallel merges}
We redo the benchmarking in Section~\ref{sec:lcs_par_merges} but with GB codes instead. We again restrict the benchmark to only parallelise up to 7 merges for efficiency of computation.

\begin{figure}
\begin{center}
\begin{tabular}{|c||c|c|c|}
\hline
$\ell$ & $r $ & $ n_{\rm ancilla}/{n_{\rm initial}} $ & $\omega $ \\
\hhline{|=||=|=|=|}
23 & 1 & 0.46 & 10 \\
\hline
24 & 1 & 1.125 & 14 \\
\hline
63 & 2 & 4.34 & 17 \\
\hline
90 & 1 & 1.34 & 15 \\
\hline
127 & 1 & 1.66 & 17 \\
\hline
\end{tabular}

\vspace{5mm}

\begin{tabular}{|c||c|c|c|}
\hline
$\ell$ & $r $ & $ n_{\rm ancilla}/{n_{\rm initial}} $ & $\omega $ \\
\hhline{|=||=|=|=|}
23 & 1 & 0.42 & 10 \\
\hline
24 & 1 & 1.35 & 14 \\
\hline
63 & 1 & 1.71 & 17 \\
\hline
90 & 1 & 1.3 & 15 \\
\hline
127 & 1 & 1.64 & 17 \\
\hline
\end{tabular}
\end{center}
\caption{Figures of merit for parallel $X$ and $Z$ merges between GB codes.}\label{fig:Z_par_merges_gb}
\end{figure}

\begin{figure}
\begin{center}
\begin{tabular}{|c||c|c|c|}
\hline
$\ell$ & $n_{\rm initial}$ & $ n_{\rm ancilla} $ & $ n_{\rm total} $ \\
\hhline{|=||=|=|=|}
23 & 92 & 39 & 131 \\
  & 580 & 16  & 596   \\
  & 92 & 1120 & 1212  \\
\hline
24 & 96 & 130 & 226 \\
   & 1356 & 42   & 1398  \\
   & 96 & 3248  & 3344  \\
\hline
63 & 252 & 430 & 682 \\
   & 6328 & 49   & 6377  \\
   & 252 & 10688 & 10940 \\
\hline
90 & 360 & 468 & 828 \\
  & 12260 & 119 & 12379 \\
  & 360 & 27004 & 27364 \\
\hline
127 & 508 & 831 & 1339 \\
   & 42616 & 133 & 42749  \\
   & 508 & 52456 & 52964  \\
\hline
\end{tabular}
\end{center}
\caption{Comparison of GB code parallel $Z$-merges to surface codes and \cite{Coh}. The first row in each box is our homological approach using Algorithm~\ref{alg:ext_merge}. The second is lattice surgery with surface code patches. The third is a naive application of \cite{Coh} to GB codes.}\label{fig:Z_GB_par_compare}
\end{figure}

We see that even when parallelised, the merges are still quite cheap; in particular the total number of qubits required increases much slower than for surface codes or \cite{Coh}. We do not have a good understanding of what about GB codes allows for this efficiency, and it is possible that \verb|QDistRnd| is giving an upper bound on distance which isn't tight, which would lead to merges which appear more efficient than they are. Lastly, the stabiliser weights increase substantially, up to a maximum of 17, to the point that fault-tolerance may be impossible with such merged codes.

\subsection{Individual single-qubit measurements}

This section is a re-run of Section~\ref{sec:lcs_ind_singleqs} but for GB codes, and once again using only the first 7 logical qubits. See Figure~\ref{fig:ind_singleqs_gb} and Figure~\ref{fig:GB_ind_singleqs_compare} for the results.

\begin{figure}
\begin{center}
\begin{tabular}{|c||c|c|c|}
\hline
$\ell$ & $\langle r\rangle$ & $\langle n_{\rm ancilla}/{n_{\rm initial}} \rangle$ & $\langle \omega \rangle$ \\
\hhline{|=||=|=|=|}
23 & 1.5 & 0.85 & 9 \\
\hline
24 & 1.7 & 0.77 & 9 \\
\hline
63 & 1.29 & 0.7 & 11 \\
\hline
90 & 1.57 & 0.74 & 9 \\
\hline
127 & 1 & 0.47 & 11 \\
\hline
\end{tabular}

\vspace{5mm}

\begin{tabular}{|c||c|c|c|}
\hline
$\ell$ & $\langle r\rangle$ & $\langle n_{\rm ancilla}/{n_{\rm initial}} \rangle$ & $\langle \omega \rangle$ \\
\hhline{|=||=|=|=|}
23 & 2 & 1.12 & 9 \\
\hline
24 & 1.5 & 0.86 & 9 \\
\hline
63 & 1 & 0.49 & 11 \\
\hline
90 & 1.71 & 0.8 & 9 \\
\hline
127 & 1 & 0.47 & 11 \\
\hline
\end{tabular}
\end{center}
\caption{Figures of merit for individual single-qubit logical $X$ and $Z$ measurements with GB codes.}\label{fig:ind_singleqs_gb}
\end{figure}

\begin{figure}
\begin{center}
\begin{tabular}{|c||c|c|c|}
\hline
$\ell$ & $n_{\rm initial}$ & $\langle n_{\rm ancilla} \rangle$ & $\langle n_{\rm total} \rangle$ \\
\hhline{|=||=|=|=|}
23 & 46 & 51.52 & 97.52 \\
  & 290 & 0  & 290   \\
  & 46 & 275.5 & 321.5  \\
\hline
24 & 48 & 41.28 & 89.28 \\
   & 678 & 0   & 678  \\
   & 48 & 266.67  & 314.67  \\
\hline
63 & 126 & 61.74 & 187.74 \\
   & 3164 & 0   & 3164  \\
   & 126 & 759.43 & 885.42 \\
\hline
90 & 180 & 144 & 324 \\
  & 6130 & 0 & 6130 \\
  & 180 & 1919.86 & 2099.86 \\
\hline
127 & 254 & 119.38 & 373.38 \\
   & 42616 & 133 & 42749  \\
   & 254 & 3736.86 & 3990.86  \\
\hline
\end{tabular}
\end{center}
\caption{Comparison of GB code individual single-qubit logical $Z$-measurements to surface codes and a naive application of \cite{Coh}. The first row uses the method described in Section~\ref{sec:single_qubit}. The second is lattice surgery with surface code patches. The third is a naive application of \cite{Coh} to GB codes.}\label{fig:GB_ind_singleqs_compare}
\end{figure}

\subsection{Parallel single-qubit measurements}

This time we re-run Section~\ref{sec:lcs_par_singleqs}, but with the same modifications for GB codes: we use \verb|QDistRnd| to estimate code distances, and measure at most the first 7 logical qubits in parallel. See Figure~\ref{fig:par_singleqs_gb} and Figure~\ref{fig:GB_par_singleqs_compare} for results. Unfortunately, the compute time of parallel single-qubit measurements for $\ell = 127$ became too high, so we omit this data.

\begin{figure}
\begin{center}
\begin{tabular}{|c||c|c|c|}
\hline
$\ell$ & $ r$ & $ n_{\rm ancilla}/{n_{\rm initial}} $ & $\omega $ \\
\hhline{|=||=|=|=|}
23 & 2 & 1.26 & 9 \\
\hline
24 & 2 & 3 & 11 \\
\hline
63 & 2 & 8.68 & 17 \\
\hline
90 & 2 & 5.46 & 13 \\
\hline
\end{tabular}

\vspace{5mm}

\begin{tabular}{|c||c|c|c|}
\hline
$\ell$ & $ r$ & $ n_{\rm ancilla}/{n_{\rm initial}} $ & $ \omega$ \\
\hhline{|=||=|=|=|}
23 & 2 & 1.15 & 9 \\
\hline
24 & 3 & 6.15 & 11 \\
\hline
63 & 1 & 3.41 & 17 \\
\hline
90 & 2 & 4.87 & 13 \\
\hline
\end{tabular}
\end{center}
\caption{Figures of merit for parallel single-qubit logical $X$ and $Z$ measurements with GB codes.}\label{fig:par_singleqs_gb}
\end{figure}

\begin{figure}
\begin{center}
\begin{tabular}{|c||c|c|c|}
\hline
$\ell$ & $n_{\rm initial}$ & $n_{\rm ancilla} $ & $ n_{\rm total} $ \\
\hhline{|=||=|=|=|}
23 & 46 & 53 & 99 \\
  & 290 & 0  & 290   \\
  & 46 & 551 & 597  \\
\hline
24 & 48 & 295 & 343 \\
   & 678 & 0   & 678  \\
   & 48 & 1600  & 1648  \\
\hline
63 & 126 & 430 & 556 \\
   & 3164 & 0   & 3164  \\
   & 126 & 5316 & 5442 \\
\hline
90 & 180 & 877 & 1057 \\
  & 6130 & 0 & 6130 \\
  & 180 & 13439 & 13619 \\
\hline
\end{tabular}
\end{center}
\caption{Comparison of GB code parallel single-qubit logical $Z$-measurements to surface codes and a naive application of \cite{Coh}. The first row uses the method described in Section~\ref{sec:single_qubit}. The second is lattice surgery with surface code patches. The third is a naive application of \cite{Coh} to GB codes.}\label{fig:GB_par_singleqs_compare}
\end{figure}

\section{Detailed results}\label{app:more_results}
In this appendix we present more fine-grained versions of the tables presented in Section~\ref{sec:auto_external}. Throughout, $i$ refers to the logical qubit used in the merge/measurement.

\begin{figure}
\begin{center}
\begin{tabular}{|c|c|c||c|c|c|c|}
\hline
$L$ & $\ell$ & $i$ & $r$ & $n_{\rm ancilla}/{n_{\rm initial}}$ & $\omega$ \\
\hhline{|=|=|=||=|=|=|}
1 & 3 & 0 & 1 & 0.17 & 6 \\
\hline
1 & 3 & 1 & 1 & 0.17 & 6 \\
\hline
1 & 3 & 2 & 1 & 0.1 & 6 \\
\hline
1 & 4 & 0 & 1 & 0.125 & 6 \\
\hline
1 & 4 & 1 & 1 & 0.175 & 6 \\
\hline
1 & 4 & 2 & 1 & 0.125 & 6 \\
\hline
1 & 4 & 3 & 1 & 0.075 & 6 \\
\hline
1 & 5 & 0 & 1 & 0.1 & 6 \\
\hline
1 & 5 & 1 & 1 & 0.16 & 6 \\
\hline
1 & 5 & 2 & 1 & 0.14 & 6 \\
\hline
1 & 5 & 3 & 1 & 0.1 & 6 \\
\hline
1 & 5 & 4 & 1 & 0.06 & 6 \\
\hline
2 & 4 & 0 & 2 & 0.25 & 7 \\
\hline
2 & 4 & 1 & 2 & 0.36 & 7 \\
\hline
2 & 4 & 2 & 2 & 0.25 & 7 \\
\hline
2 & 4 & 3 & 1 & 0.11 & 7 \\
\hline
2 & 5 & 0 & 1 & 0.069 & 7 \\
\hline
2 & 5 & 1 & 1 & 0.1 & 7 \\
\hline
2 & 5 & 2 & 1 & 0.092 & 7 \\
\hline
2 & 5 & 3 & 1 & 0.069 & 7 \\
\hline
2 & 5 & 4 & 3 & 0.49 & 7 \\
\hline
2 & 6 & 0 & 1 & 0.058 & 7 \\
\hline
2 & 6 & 1 & 1 & 0.09 & 7 \\
\hline
2 & 6 & 2 & 1 & 0.071 & 7 \\
\hline
2 & 6 & 3 & 1 & 0.077 & 7 \\
\hline
2 & 6 & 4 & 1 & 0.058 & 7 \\
\hline
2 & 6 & 5 & 3 & 0.49 & 7 \\
\hline
3 & 5 & 0 & 1 & 0.052 & 7 \\
\hline
3 & 5 & 1 & 1 & 0.048 & 7 \\
\hline
3 & 5 & 2 & 1 & 0.072 & 7 \\
\hline
3 & 5 & 3 & 1 & 0.096 & 7 \\
\hline
3 & 5 & 4 & 3 & 0.456 & 7 \\
\hline
3 & 6 & 0 & 2 & 0.11 & 7 \\
\hline
3 & 6 & 1 & 1 & 0.06 & 7 \\
\hline
3 & 6 & 2 & 1 & 0.057 & 7 \\
\hline
3 & 6 & 3 & 2 & 0.12 & 7 \\
\hline
3 & 6 & 4 & 2 & 0.23 & 7 \\
\hline
3 & 6 & 5 & 2 & 0.3 & 7 \\
\hline
\end{tabular}
\end{center}
\caption{Expanded figures of merit for individual $Z$-merges between LCS codes.}\label{fig:expnd_Z_ind_merges_lcs}
\end{figure}

\begin{figure}
\begin{center}
\begin{tabular}{|c|c|c||c|c|c|c|}
\hline
$L$ & $\ell$ & $i$ & $r$ & $n_{\rm ancilla}/{n_{\rm initial}}$ & $\omega$ \\
\hhline{|=|=|=||=|=|=|}
1 & 3 & 0 & 1 & 0.17 & 6 \\
\hline
1 & 3 & 1 & 1 & 0.2 & 6 \\
\hline
1 & 3 & 2 & 1 & 0.1 & 6 \\
\hline
1 & 4 & 0 & 1 & 0.15 & 6 \\
\hline
1 & 4 & 1 & 1 & 0.125 & 6 \\
\hline
1 & 4 & 2 & 1 & 0.2 & 6 \\
\hline
1 & 4 & 3 & 1 & 0.075 & 6 \\
\hline
1 & 5 & 0 & 1 & 0.14 & 6 \\
\hline
1 & 5 & 1 & 1 & 0.12 & 6 \\
\hline
1 & 5 & 2 & 1 & 0.1 & 6 \\
\hline
1 & 5 & 3 & 1 & 0.2 & 6 \\
\hline
1 & 5 & 4 & 1 & 0.06 & 6 \\
\hline
2 & 4 & 0 & 1 & 0.096 & 7 \\
\hline
2 & 4 & 1 & 1 & 0.14 & 7 \\
\hline
2 & 4 & 2 & 1 & 0.15 & 7 \\
\hline
2 & 4 & 3 & 1 & 0.048 & 7 \\
\hline
2 & 5 & 0 & 2 & 0.28 & 7 \\
\hline
2 & 5 & 1 & 2 & 0.2 & 7 \\
\hline
2 & 5 & 2 & 2 & 0.37 & 7 \\
\hline
2 & 5 & 3 & 3 & 0.91 & 7 \\
\hline
2 & 5 & 4 & 1 & 0.038 & 7 \\
\hline
2 & 6 & 0 & 2 & 0.26 & 7 \\
\hline
2 & 6 & 1 & 2 & 0.24 & 7 \\
\hline
2 & 6 & 2 & 2 & 0.19 & 7 \\
\hline
2 & 6 & 3 & 2 & 0.35 & 7 \\
\hline
2 & 6 & 4 & 3 & 0.74 & 7 \\
\hline
2 & 6 & 5 & 1 & 0.032 & 7 \\
\hline
3 & 5 & 0 & 2 & 0.21 & 7 \\
\hline
3 & 5 & 1 & 2 & 0.3 & 7 \\
\hline
3 & 5 & 2 & 3 & 0.5 & 7 \\
\hline
3 & 5 & 3 & 3 & 0.71 & 7 \\
\hline
3 & 5 & 4 & 3 & 0.15 & 7 \\
\hline
3 & 6 & 0 & 2 & 0.19 & 7 \\
\hline
3 & 6 & 1 & 4 & 0.40 & 7 \\
\hline
3 & 6 & 2 & 2 & 0.30 & 7 \\
\hline
3 & 6 & 3 & 3 & 0.46 & 7 \\
\hline
3 & 6 & 4 & 2 & 0.35 & 7 \\
\hline
3 & 6 & 5 & 4 & 0.18 & 7 \\
\hline
\end{tabular}
\end{center}
\caption{Expanded figures of merit for individual $X$-merges between LCS codes.}\label{fig:expnd_X_ind_merges_lcs}
\end{figure}

\begin{figure}
\begin{center}
\begin{tabular}{|c|c|c||c|c|c|c|}
\hline
$L$ & $\ell$ & $i$ & $r$ & $n_{\rm ancilla}/{n_{\rm initial}}$ & $\omega$ \\
\hhline{|=|=|=||=|=|=|}
1 & 3 & 0 & 2 & 1.1 & 6 \\
\hline
1 & 3 & 1 & 2 & 1.1 & 6 \\
\hline
1 & 3 & 2 & 1 & 0.2 & 6 \\
\hline
1 & 4 & 0 & 2 & 0.8 & 6 \\
\hline
1 & 4 & 1 & 2 & 1.05 & 6 \\
\hline
1 & 4 & 2 & 2 & 0.8 & 6 \\
\hline
1 & 4 & 3 & 1 & 0.15 & 6 \\
\hline
1 & 5 & 0 & 2 & 0.64 & 6 \\
\hline
1 & 5 & 1 & 2 & 0.96 & 6 \\
\hline
1 & 5 & 2 & 2 & 0.84 & 6 \\
\hline
1 & 5 & 3 & 2 & 0.64 & 6 \\
\hline
1 & 5 & 4 & 1 & 0.12 & 6 \\
\hline
2 & 4 & 0 & 1 & 0.17 & 7 \\
\hline
2 & 4 & 1 & 1 & 0.25 & 7 \\
\hline
2 & 4 & 2 & 1 & 0.17 & 7 \\
\hline
2 & 4 & 3 & 2 & 0.6 & 7 \\
\hline
2 & 5 & 0 & 2 & 0.4 & 7 \\
\hline
2 & 5 & 1 & 2 & 0.58 & 7 \\
\hline
2 & 5 & 2 & 2 & 0.54 & 7 \\
\hline
2 & 5 & 3 & 2 & 0.4 & 7 \\
\hline
2 & 5 & 4 & 3 & 0.98 & 7 \\
\hline
2 & 6 & 0 & 2 & 0.33 & 7 \\
\hline
2 & 6 & 1 & 2 & 0.53 & 7 \\
\hline
2 & 6 & 2 & 2 & 0.41 & 7 \\
\hline
2 & 6 & 3 & 2 & 0.45 & 7 \\
\hline
2 & 6 & 4 & 2 & 0.33 & 7 \\
\hline
2 & 6 & 5 & 3 & 0.99 & 7 \\
\hline
3 & 5 & 0 & 1 & 0.11 & 7 \\
\hline
3 & 5 & 1 & 2 & 0.28 & 7 \\
\hline
3 & 5 & 2 & 2 & 0.42 & 7 \\
\hline
3 & 5 & 3 & 2 & 0.54 & 7 \\
\hline
3 & 5 & 4 & 4 & 1.28 & 7 \\
\hline
3 & 6 & 0 & 2 & 0.21 & 7 \\
\hline
3 & 6 & 1 & 3 & 0.59 & 7 \\
\hline
3 & 6 & 2 & 1 & 0.11 & 7 \\
\hline
3 & 6 & 3 & 3 & 0.39 & 7 \\
\hline
3 & 6 & 4 & 3 & 0.77 & 7 \\
\hline
3 & 6 & 5 & 4 & 1.4 & 7 \\
\hline
\end{tabular}
\end{center}
\caption{Expanded figures of merit for individual single qubit logical $Z$-measurements with LCS codes.}\label{fig:expnd_Z_singleqs_lcs}
\end{figure}

\begin{figure}
\begin{center}
\begin{tabular}{|c|c|c||c|c|c|}
\hline
$L$ & $\ell$ & $i$ & $r$ & $n_{\rm ancilla}/{n_{\rm initial}}$ & $\omega$ \\
\hhline{|=|=|=||=|=|=|}
1 & 3 & 0 & 2 & 1.0 & 6 \\
\hline
1 & 3 & 1 & 2 & 1.2 & 6 \\
\hline
1 & 3 & 2 & 1 & 0.2 & 6 \\
\hline
1 & 4 & 0 & 2 & 0.9 & 6 \\
\hline
1 & 4 & 1 & 2 & 0.75 & 6 \\
\hline
1 & 4 & 2 & 2 & 1.2 & 6 \\
\hline
1 & 4 & 3 & 1 & 0.15 & 6 \\
\hline
1 & 5 & 0 & 2 & 0.84 & 6 \\
\hline
1 & 5 & 1 & 2 & 0.72 & 6 \\
\hline
1 & 5 & 2 & 2 & 0.6 & 6 \\
\hline
1 & 5 & 3 & 2 & 1.2 & 6 \\
\hline
1 & 5 & 4 & 1 & 0.12 & 6 \\
\hline
2 & 4 & 0 & 1 & 0.19 & 7 \\
\hline
2 & 4 & 1 & 2 & 0.81 & 7 \\
\hline
2 & 4 & 2 & 2 & 0.88 & 7 \\
\hline
2 & 4 & 3 & 1 & 0.1 & 7 \\
\hline
2 & 5 & 0 & 3 & 0.94 & 7 \\
\hline
2 & 5 & 1 & 3 & 0.66 & 7 \\
\hline
2 & 5 & 2 & 4 & 1.69 & 7 \\
\hline
2 & 5 & 3 & 4 & 2.54 & 7 \\
\hline
2 & 5 & 4 & 1 & 0.08 & 7 \\
\hline
2 & 6 & 0 & 3 & 0.85 & 7 \\
\hline
2 & 6 & 1 & 3 & 0.78 & 7 \\
\hline
2 & 6 & 2 & 2 & 0.37 & 7 \\
\hline
2 & 6 & 3 & 3 & 1.14 & 7 \\
\hline
2 & 6 & 4 & 3 & 1.49 & 7 \\
\hline
2 & 6 & 5 & 1 & 0.06 & 7 \\
\hline
3 & 5 & 0 & 3 & 0.7 & 7 \\
\hline
3 & 5 & 1 & 3 & 1.0 & 7 \\
\hline
3 & 5 & 2 & 3 & 1.01 & 7 \\
\hline
3 & 5 & 3 & 3 & 1.42 & 7 \\
\hline
3 & 5 & 4 & 1 & 0.06 & 7 \\
\hline
3 & 6 & 0 & 5 & 1.09 & 7 \\
\hline
3 & 6 & 1 & 4 & 0.81 & 7 \\
\hline
3 & 6 & 2 & 3 & 1.0 & 7 \\
\hline
3 & 6 & 3 & 4 & 1.29 & 7 \\
\hline
3 & 6 & 4 & 4 & 1.6 & 7 \\
\hline
3 & 6 & 5 & 1 & 0.05 & 7 \\
\hline
\end{tabular}
\end{center}
\caption{Expanded figures of merit for individual single qubit logical $X$-measurements with LCS codes.}\label{fig:expnd_X_singleqs_lcs}
\end{figure}

\begin{figure}
\begin{center}
\begin{tabular}{|c|c||c|c|c|}
\hline
$\ell$ & $i$ & $r$ & $n_{\rm ancilla}/{n_{\rm initial}}$ & $\omega$ \\
\hhline{|=|=||=|=|=|}
23 & 0 & 1 & 0.22 & 9 \\
\hline
23 & 1 & 1 & 0.21 & 9 \\
\hline
24 & 0 & 2 & 0.69 & 9 \\
\hline
24 & 1 & 1 & 0.22 & 9 \\
\hline
24 & 2 & 1 & 0.25 & 9 \\
\hline
24 & 3 & 1 & 0.23 & 9 \\
\hline
24 & 4 & 1 & 0.21 & 9 \\
\hline
24 & 5 & 1 & 0.2 & 9  \\
\hline
63 & 0 & 1 & 0.25 & 11 \\
\hline
63 & 1 & 1 & 0.24 & 11 \\
\hline
63 & 2 & 1 & 0.25 & 11  \\
\hline
63 & 3 & 1 & 0.24 & 11 \\
\hline
63 & 4 & 1 & 0.25 & 11 \\
\hline
63 & 5 & 1 & 0.25 & 11 \\
\hline
63 & 6 & 1 & 0.24 & 11 \\
\hline
90 & 0 & 1 & 0.18 & 9 \\
\hline
90 & 1 & 1 & 0.19 & 9 \\
\hline
90 & 2 & 1 & 0.17 & 9 \\
\hline
90 & 3 & 1 & 0.21 & 9 \\
\hline
90 & 4 & 1 & 0.18 & 9 \\
\hline
90 & 5 & 1 & 0.19 & 9 \\
\hline
90 & 6 & 1 & 0.18 & 9 \\
\hline
127 & 0 & 1 & 0.22 & 11 \\
\hline
127 & 1 & 1 & 0.25 & 11 \\
\hline
127 & 2 & 1 & 0.23 & 11 \\
\hline
127 & 3 & 1 & 0.23 & 11 \\
\hline
127 & 4 & 1 & 0.23 & 11 \\
\hline
127 & 5 & 1 & 0.23 & 11 \\
\hline
127 & 6 & 1 & 0.24 & 11 \\
\hline
\end{tabular}
\end{center}
\caption{Expanded figures of merit for individual $Z$-merges between GB codes.}\label{fig:expnd_Z_ind_merges_gb}
\end{figure}

\begin{figure}
\begin{center}
\begin{tabular}{|c|c||c|c|c|}
\hline
$\ell$ & $i$ & $r$ & $n_{\rm ancilla}/{n_{\rm initial}}$ & $\omega$ \\
\hhline{|=|=||=|=|=|}
23 & 0 & 1 & 0.24 & 9 \\
\hline
23 & 1 & 1 & 0.22 & 9 \\
\hline
24 & 0 & 1 & 0.2 & 9 \\
\hline
24 & 1 & 1 & 0.2 & 9 \\
\hline
24 & 2 & 1 & 0.2 & 9 \\
\hline
24 & 3 & 1 & 0.2 & 9 \\
\hline
24 & 4 & 1 & 0.17 & 9 \\
\hline
24 & 5 & 1 & 0.17 & 9 \\
\hline
63 & 0 & 1 & 0.23 & 11 \\
\hline
63 & 1 & 1 & 0.23 & 11 \\
\hline
63 & 2 & 1 & 0.25 & 11 \\
\hline
63 & 3 & 1 & 0.25 & 11 \\
\hline
63 & 4 & 2 & 0.63 & 11 \\
\hline
63 & 5 & 1 & 0.23 & 11 \\
\hline
63 & 6 & 2 & 0.63 & 11 \\
\hline
90 & 0 & 1 & 0.22 & 9 \\
\hline
90 & 1 & 2 & 0.56 & 9 \\
\hline
90 & 2 & 1 & 0.21 & 9 \\
\hline
90 & 3 & 1 & 0.18 & 9 \\
\hline
90 & 4 & 1 & 0.22 & 9 \\
\hline
90 & 5 & 1 & 0.16 & 9 \\
\hline
90 & 6 & 1 & 0.16 & 9 \\
\hline
127 & 0 & 1 & 0.23 & 11 \\
\hline
127 & 1 & 1 & 0.24 & 11 \\
\hline
127 & 2 & 1 & 0.24 & 11 \\
\hline
127 & 3 & 1 & 0.24 & 11 \\
\hline
127 & 4 & 1 & 0.22 & 11 \\
\hline
127 & 5 & 1 & 0.24 & 11 \\
\hline
127 & 6 & 1 & 0.24 & 11 \\
\hline
\end{tabular}
\end{center}
\caption{Expanded figures of merit for individual $X$-merges between GB codes.}\label{fig:expnd_X_ind_merges_gb}
\end{figure}

\begin{figure}
\begin{center}
\begin{tabular}{|c|c||c|c|c|}
\hline
$\ell$ & $i$ & $r$ & $n_{\rm ancilla}/{n_{\rm initial}}$ & $\omega$ \\
\hhline{|=|=||=|=|=|}
23 & 0 & 2 & 1.15 & 9 \\
\hline
23 & 1 & 2 & 1.09 & 9 \\
\hline
24 & 0 & 3 & 2.25 & 9 \\
\hline
24 & 1 & 1 & 0.44 & 9 \\
\hline
24 & 2 & 1 & 0.5 & 9 \\
\hline
24 & 3 & 1 & 0.46 & 9 \\
\hline
24 & 4 & 2 & 1.08 & 9 \\
\hline
24 & 5 & 1 & 0.4 & 9 \\
\hline
63 & 0 & 1 & 0.5 & 11 \\
\hline
63 & 1 & 1 & 0.48 & 11 \\
\hline
63 & 2 & 1 & 0.5 & 11 \\
\hline
63 & 3 & 1 & 0.49 & 11 \\
\hline
63 & 4 & 1 & 0.49 & 11 \\
\hline
63 & 5 & 1 & 0.49 & 11 \\
\hline
63 & 6 & 1 & 0.48 & 11 \\
\hline
90 & 0 & 2 & 0.94 & 9 \\
\hline
90 & 1 & 1 & 0.38 & 9 \\
\hline
90 & 2 & 2 & 0.89 & 9 \\
\hline
90 & 3 & 2 & 1.08 & 9 \\
\hline
90 & 4 & 2 & 0.97 & 9 \\
\hline
90 & 5 & 2 & 1.0 & 9 \\
\hline
90 & 6 & 1 & 0.37 & 9 \\
\hline
127 & 0 & 1 & 0.43 & 11 \\
\hline
127 & 1 & 1 & 0.49 & 11 \\
\hline
127 & 2 & 1 & 0.46 & 11 \\
\hline
127 & 3 & 1 & 0.47 & 11 \\
\hline
127 & 4 & 1 & 0.46 & 11 \\
\hline
127 & 5 & 1 & 0.47 & 11 \\
\hline
127 & 6 & 1 & 0.48 & 11 \\
\hline
\end{tabular}
\end{center}
\caption{Expanded figures of merit for individual single qubit logical $Z$-measurements with GB codes.}\label{fig:expnd_Z_singleqs_gb}
\end{figure}

\begin{figure}
\begin{center}
\begin{tabular}{|c|c||c|c|c|}
\hline
$\ell$ & $i$ & $r$ & $n_{\rm ancilla}/{n_{\rm initial}}$ & $\omega$ \\
\hhline{|=|=||=|=|=|}
23 & 0 & 2 & 1.26 & 9 \\
\hline
23 & 1 & 1 & 0.43 & 9 \\
\hline
24 & 0 & 2 & 1.0 & 9 \\
\hline
24 & 1 & 2 & 1.0 & 9 \\
\hline
24 & 2 & 2 & 1.0 & 9 \\
\hline
24 & 3 & 1 & 0.4 & 9 \\
\hline
24 & 4 & 1 & 0.33 & 9 \\
\hline
24 & 5 & 2 & 0.88 & 9 \\
\hline
63 & 0 & 1 & 0.47 & 11 \\
\hline
63 & 1 & 1 & 0.47 & 11 \\
\hline
63 & 2 & 1 & 0.5 & 11 \\
\hline
63 & 3 & 1 & 0.49 & 11 \\
\hline
63 & 4 & 2 & 1.25 & 11 \\
\hline
63 & 5 & 1 & 0.47 & 11 \\
\hline
63 & 6 & 2 & 1.25 & 11 \\
\hline
90 & 0 & 1 & 0.43 & 9 \\
\hline
90 & 1 & 2 & 1.12 & 9 \\
\hline
90 & 2 & 1 & 0.42 & 9 \\
\hline
90 & 3 & 2 & 0.94 & 9 \\
\hline
90 & 4 & 2 & 1.17 & 9 \\
\hline
90 & 5 & 2 & 0.82 & 9 \\
\hline
90 & 6 & 1 & 0.31 & 9 \\
\hline
127 & 0 & 1 & 0.46 & 11 \\
\hline
127 & 1 & 1 & 0.48 & 11 \\
\hline
127 & 2 & 1 & 0.49 & 11 \\
\hline
127 & 3 & 1 & 0.48 & 11 \\
\hline
127 & 4 & 1 & 0.45 & 11 \\
\hline
127 & 5 & 1 & 0.48 & 11 \\
\hline
127 & 6 & 1 & 0.47 & 11 \\
\hline
\end{tabular}
\end{center}
\caption{Expanded figures of merit for individual single qubit logical $X$-measurements with GB codes.}\label{fig:expnd_X_singleqs_gb}
\end{figure}